\documentclass[prx,twocolumn,floatfix,superscriptaddress, longbibliography]{revtex4-1}
\usepackage[pdftex,plainpages=false,colorlinks=true,linkcolor=blue, citecolor=blue, urlcolor=blue]{hyperref}
\usepackage{amsfonts}
\usepackage{amsmath}
\usepackage{amssymb}
\usepackage{graphicx}
\usepackage{natbib}
\usepackage{color}
\usepackage{placeins}
\usepackage{physics}

\begin{document}

\title{Information-theoretic measures of superconductivity in a two-dimensional doped Mott insulator}

\author{C. Walsh}
\affiliation{Department of Physics, Royal Holloway, University of London, Egham, Surrey, UK, TW20 0EX}
\author{M. Charlebois}
\affiliation{D\'epartement de Chimie, Biochimie et Physique, Institut de Recherche sur l’Hydrog\`ene, Universit\'e du Qu\'ebec \`a Trois-Rivi\`eres, Trois-Rivi\`eres, Qu\'ebec, Canada G9A 5H7}
\author{P. S\'emon}
\affiliation{Computational Science Initiative, Brookhaven National Laboratory, Upton, NY 11973-5000, USA}
\author{G. Sordi}
\email[corresponding author: ]{giovanni.sordi@rhul.ac.uk}
\affiliation{Department of Physics, Royal Holloway, University of London, Egham, Surrey, UK, TW20 0EX}
\author{A.-M. S. Tremblay}
\affiliation{D\'epartement de physique \& Institut quantique, Universit\'e de Sherbrooke, Sherbrooke, Qu\'ebec, Canada J1K 2R1}

\date{\today}

\begin{abstract}
A key open issue in condensed matter physics is how quantum and classical correlations emerge in an unconventional superconductor from the underlying normal state. We study this problem in a doped Mott insulator with information theory tools on the two-dimensional Hubbard model at finite temperature with cluster dynamical mean-field theory. We find that the local entropy detects the superconducting state and that the difference in the local entropy between the superconducting and normal states follows the same difference in the potential energy. We find that the thermodynamic entropy is suppressed in the superconducting state and monotonically decreases with decreasing doping. The maximum in entropy found in the normal state above the overdoped region of the superconducting dome is obliterated by superconductivity. The total mutual information, which quantifies quantum and classical correlations, is amplified in the superconducting state of the doped Mott insulator for all doping levels, and shows a broad peak versus doping, as a result of competing quantum and classical effects.
\end{abstract}

\maketitle



\section{Introduction}
Quantum and classical correlations among electrons in many-body quantum systems give rise to striking phases of matter. Understanding the nature of these correlations is a fundamental open problem in many-body quantum physics~\cite{zhengBOOK}. A prominent example is the elusive mechanism of high-temperature superconductivity in doped Mott insulators, as realised in cuprates~\cite{Anderson:1987, lee, tremblayR}. 

The two-dimensional Hubbard model is the simplest theoretical framework for the phenomenology of cuprates~\cite{Alloul2013}, and can be realised with ultracold atoms in optical lattices~\cite{jzHM, Esslinger:2010, GrossScience2017}. 
The {\it thermodynamics} of the superconducting state and its underlying normal state have been intensively investigated within this model, with good agreement with experiments in cuprates~\cite{Alloul2013}. In contrast, there is little knowledge about how {\it quantum information concepts} based on entanglement-related properties characterise the superconducting state~\cite{amicoRMP2008}. Understanding this is crucial, as entanglement may provide a more fundamental description of quantum matter~\cite{amicoRMP2008, zhengBOOK, Laflorencie:PhysRep2016}. Furthermore, lattices of ultracold atoms can now detect entanglement-related quantities~\cite{greinerNat2015, Kaufman:Science2016, Cocchi:PRX2017, Lukin:Science2019} and are close to being able to detect superconducting correlations~\cite{GrossScience2017, Hofstetter:2002}. 

In this article we examine how information-theoretic measures can be used to provide an improved understanding of strongly correlated superconductivity. Entropy is a key concept at the foundation of information theory~\cite{Jaynes:1957, watrous2018} and thus quantities based on entropy can be used to describe quantum and classical correlations imprinted in the superconducting state. We answer core questions: How does superconductivity quench the large thermodynamic entropy of the normal state at optimal doping? How do entanglement-related quantities - local entropy and total mutual information - detect the superconducting state? The former contains the entanglement between a site and its environment, whereas the latter measures classical and quantum correlations between a site and its environment. 
Our analysis builds upon our previous work focused on the normal state~\cite{Caitlin:PRXQ2020} and extends it to the superconducting state.

\section{Model and method}
To address these questions, we consider the two-dimensional Hubbard model on a square lattice, $H=-\sum_{\langle ij\rangle \sigma}t_{ij}c_{i\sigma}^\dagger c_{j\sigma}
+U\sum_{i} n_{i\uparrow} n_{i\downarrow}
-\mu\sum_{i\sigma} n_{i\sigma}$, where $c_{i\sigma}$ and $c^\dagger_{i\sigma}$ respectively destroy and create an electron with spin $\sigma$ at site $i$, $n=c_{i\sigma}^\dagger c_{i\sigma}$, $t_{ij}$ is the nearest neighbor hopping, $\mu$ is the chemical potential, and $U$ is the on-site Coulomb repulsion. $t_{ij}=t=1$ is our energy unit. 

We solve this model with the cellular extension~\cite{maier, kotliarRMP, tremblayR} of dynamical mean-field theory~\cite{rmp} (CDMFT). This technique isolates a cluster of lattice sites and embeds it in a self-consistent bath of non-interacting electrons. It offers the key advantage of enabling the treating of both $d$-wave superconductivity and Mott physics. We solve the cluster quantum impurity model with the hybridisation expansion continuous time quantum Monte Carlo method~\cite{millisRMP, patrickSkipList} and using Monte Carlo updates with two pairs of creation and destruction operators to ensure ergodicity in the $d$-wave superconducting state~\cite{patrickERG}. This approach allows us to study superconductivity across a wide range of interaction strengths and to attain low temperatures. Here we consider only the minimal cluster to capture $d$-wave superconductivity, i.e. the $2 \times 2$ plaquette, and to reduce the sign problem we only consider nearest neighbor hopping.

\begin{figure}[t!]
\centering{
\includegraphics[width=0.98\linewidth]{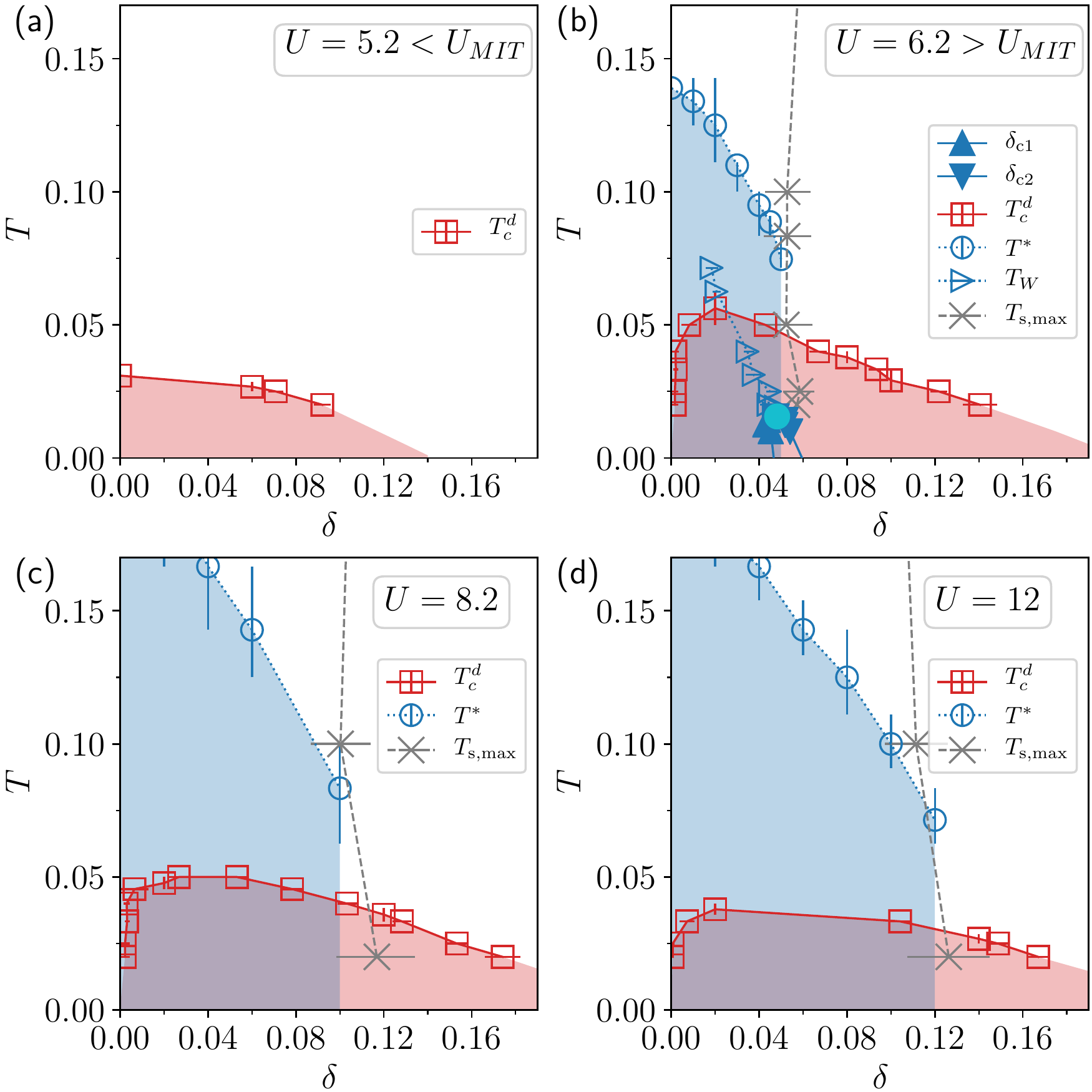}
}
\caption{Temperature $T$ versus doping $\delta$ phase diagram for (a) $U=5.2<U_{\rm MIT}$, (b) $U=6.2>U_{\rm MIT}$, (c) $U=8.2$, (d) $U=12$. 
The line of open red squares denotes the dynamical mean-field $d_{x^2-y^2}$ superconducting transition temperature $T_c^d$. 
The shaded red region underneath is a guide to the eye showing the superconducting dome.
The dotted line of blue circles shows the onset of the pseudogap $T^*$, calculated by the drop in the spin susceptibility as a function of $T$, and the shaded blue region is a guide to the eye indicating the pseudogap. At low $T$, the unshaded area indicates a correlated metallic state.
The dashed line with gray crosses shows the loci of the maximum in normal-state thermodynamic entropy $s$ as a function of doping at different temperatures, $T_{\rm s, max}$. In panel (b), the pseudogap to metal transition is shown concealed beneath the superconducting dome. The coexistence region of the first order transition is bounded by spinodal lines (filled triangles). The transition culminates at a second order critical point (filled cyan circle). Extending from the finite temperature critical point is a sharp crossover known as the Widom line that is defined here by the loci of the maximum in the isothermal charge compressibility (open triangles). Normal state data for $U=6.2$ are extracted from Ref.~\cite{Caitlin:PRXQ2020}.
}
\label{fig1}
\end{figure}

\section{Results}

\subsection{Phase diagram}
First we create a scan of the temperature $T$ versus doping $\delta$ phase diagram of the two-dimensional Hubbard model for different values of $U$, both in the superconducting and normal states, to map out the superconducting state and the underlying normal state. Fig.~\ref{fig1} shows the temperature-doping phase diagram for different values of the interaction strength $U$. The case $U=5.2$ is below the threshold for the Mott transition at half filling, which occurs at $U_{\rm MIT} \approx 5.95$~\cite{CaitlinSb}, whereas $U=6.2, 8.2, 12$ are above $U_{\rm MIT}$. 

The superconducting transition temperature $T_c^d$ (red line with squares) defines the region below which the $d_{x^2-y^2}$ superconducting order parameter $\Phi=\langle c^\dagger_{{\bf K} \uparrow} c^\dagger_{-{\bf K} \downarrow}\rangle$, with cluster momentum ${\bf K}=(\pi,0)$, is nonzero (see SI Appendix Fig.~S1). It is obtained by calculating $\Phi$ along constant-temperature  pathways and constant-doping pathways. The superscript $d$ in $T_c^d$ reminds us that $T_c^d$ is the cluster {\it dynamical} mean-field transition temperature, and thus physically indicates when superconducting pairs develop in the cluster~\cite{sshtSC}. Thermal fluctuations preclude long range order in two dimensions~\cite{MWtheorem}. However, the Kosterlitz-Thouless vortex-binding mechanism allows an onset of algebraic decay of superconducting correlations at an actual value $T_c$ less than the mean-field result~\cite{Kosterlitz:1973}. 

In agreement with Ref.~\cite{LorenzoSC}, for $U<U_{\rm MIT}$, $T_c^d$ monotonically decreases with increasing doping. In contrast, for $U>U_{\rm MIT}$, $T_c^d$ has a dome-like shape, with a maximum $(T_c^d)_{\rm max}$ occurring in the $U-\delta$ space just above $U_{\rm MIT}$. For a given $U>U_{\rm MIT}$, the superconducting dome is asymmetric in $\delta$ with a steep increase upon doping the Mott insulator, and a more gentle decrease in the overdoped region. 

Superconductivity can condense either from a strongly correlated pseudogap (blue region in Fig.~\ref{fig1}) or from a correlated metal without pseudogap. This is because the critical doping at which the pseudogap onset temperature $T^*(\delta)$ ends is in the middle of the superconducting dome, and hence superconductivity is accessible from either side~\cite{sshtSC}. 
The onset of the pseudogap $T^*(\delta)$ can be obtained by the drop in the spin susceptibility versus $T$~\cite{sshtRHO} (open circles in Fig.~\ref{fig1}). Previous studies show that $T^*(\delta)$ ends abruptly at a critical doping and temperature~\cite{sshtRHO}, as found in cuprates~\cite{Olivier:PRB2018}. $T^*$ is a high temperature precursor of a first order transition which is hidden below the superconducting state~\cite{sshtRHO, sshtSC, LorenzoSC} (in Fig.~\ref{fig1}(b), filled triangles indicate the corresponding coexistence region for $U=6.2$). This is a metal to metal transition (more precisely, a strongly correlated pseudogap to correlated metal transition), without any symmetry breaking, only changes in the electronic density. 
It is caused by Mott physics plus short range correlations~\cite{ssht}, as it is connected in the $U-\delta$ plane to the Mott insulator to metal transition at $U_{\rm MIT}$ and half filling~\cite{sht, sht2}.
This transition culminates in a critical point (filled circle in Fig.~\ref{fig1}) beyond which supercritical crossovers emerge (the Widom line~\cite{ssht, water1}). These crossovers are the loci of anomalous peaks versus doping in the electronic charge compressibility~\cite{ssht} (dotted line with open triangles for $U=6.2$), in nonlocal density fluctuations~\cite{CaitlinOpalescence}, and in the electronic specific heat~\cite{Giovanni:PRBcv}. Thermodynamic anomalies in the normal state of cuprates~\cite{Michon:Cv2018} can be rationalised by this phenomenon. 
The critical point of the pseudogap to correlated metal transition is hidden by superconductivity, but it influences the normal state up to high temperatures and controls the superconducting pairing mechanism~\cite{LorenzoSC}. 

This work focuses on entropy-based quantities, and a key feature of the entropy landscape is a maximum in the thermodynamic entropy $s$ as a function of doping. This maximum bounds the hidden pseudogap to metal transition at high doping, as shown in Fig.~\ref{fig1} and discussed in Ref.~\cite{sht2}. 
This maximum can be found by direct inspection of $s(\delta)$~\cite{Caitlin:PRXQ2020} or, using the Maxwell relation $(dn/dT)_{\mu}=(ds/d\mu)_T$~\cite{MikelsonThermodynamics:2009, sht2}, by finding the zero of the expansion coefficient $(dn/dT)_{\mu}$. As found in cuprates~\cite{LoramJPCS2001, Tallon:2004}, this is a vertical crossover line extending up to high temperature~\cite{sht2}. Physically, it is related to the localisation-delocalisation physics of doped Mott insulators~\cite{Caitlin:PRXQ2020}. It occurs close to, but not at, the critical point of the pseudogap to correlated metal transition. Indeed, the entropy shows an inflection at the critical point of this transition~\cite{Caitlin:PRXQ2020}. We shall show that the maximum in entropy found in the normal state above the overdoped region of the superconducting dome is obliterated by superconductivity.

\begin{figure}[t!]
\centering{
\includegraphics[width=0.98\linewidth]{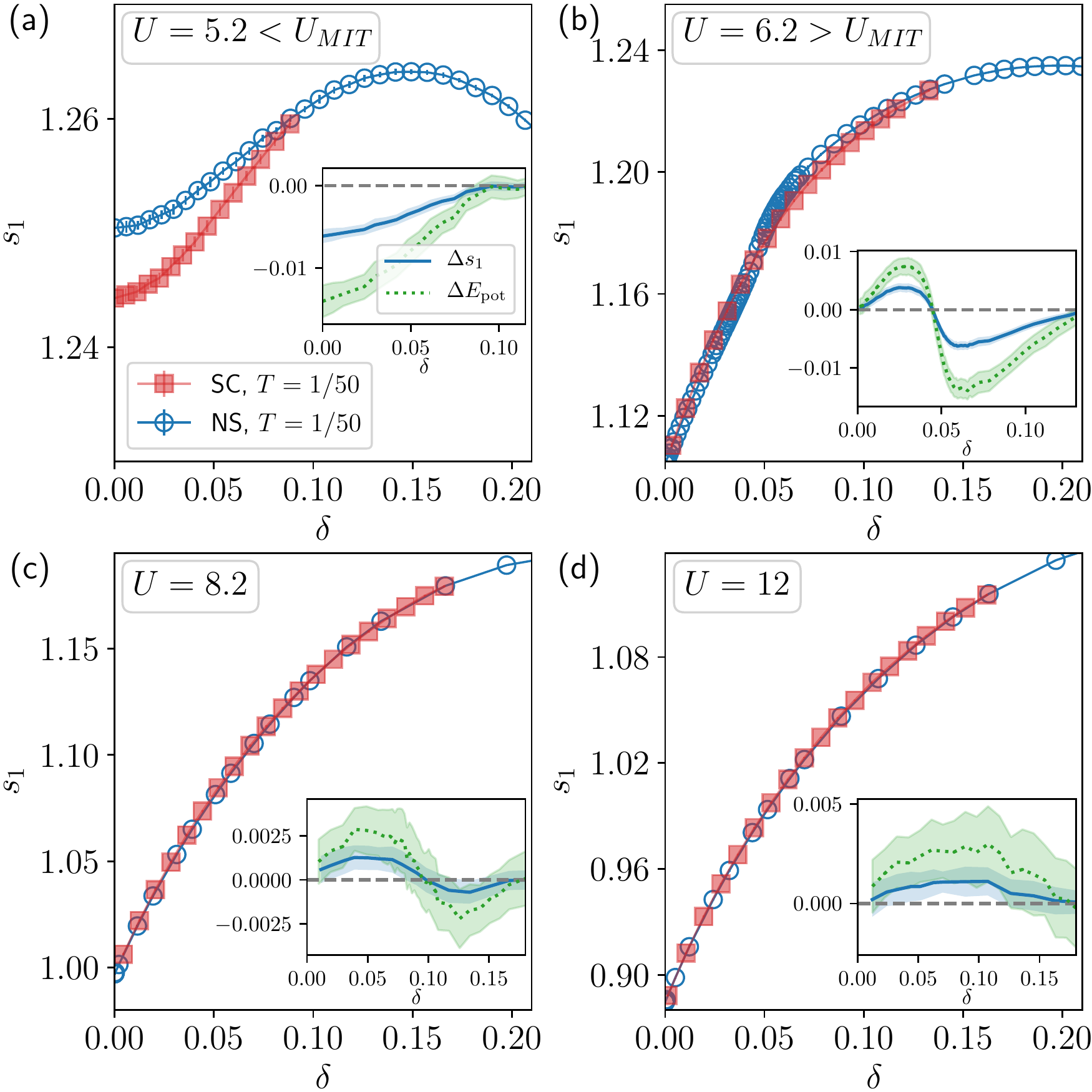}
}
\caption{Local entropy $s_1$ as a function of doping, for (a) $U=5.2<U_{\rm MIT}$, (b) $U=6.2>U_{\rm MIT}$, (c) $U=8.2$, (d) $U=12$. Data in all panels are at $T=1/50$ for the normal and superconducting states (open circles and shaded squares respectively). Insets show the difference between the normal and superconducting states for the local entropy $s_{1}$ and the potential energy $E_{\rm pot}$ as a function of doping (solid line and dotted line respectively). The shaded region around both curves shows the standard error.  
$s_1$ normal state data for $U=6.2$ are extracted from Ref.~\cite{Caitlin:PRXQ2020}.
}
\label{figS1}
\end{figure}

\subsection{Local entropy}
Next we go beyond what is accessible by a thermodynamic description and focus on information-theoretic measures of the superconducting state. We select a region of the lattice, denoting its sites with the set $A$ and all other sites by its complement $B=\overline{A}$. For a system in a pure state, entanglement properties are encoded in the reduced density matrix $\rho_A$ of subsystem $A$. In this work we let $A$ to be just a single site. In principle, measuring the entanglement properties of different sizes of the subsystem $A$ allows one to study the structure of the entanglement in the system. 

Since the system is translationally invariant and total spin $S^z=\sum_i(n_{i\uparrow}-n_{i\downarrow})$ is conserved, the reduced density matrix on one site is diagonal~\cite{zanardi2002}, and is given by $\rho_A={\rm diag}(p_0, p_{\uparrow}, p_\downarrow, p_{\uparrow\downarrow})$, where $p_i$ is the probability that the site is empty, singly occupied, and doubly occupied. This simplification allows us to extract two simple and experimentally detectable~\cite{greinerNat2015, Cocchi:PRX2017}, entanglement-related quantities, the local entropy and the total mutual information between the single site and the rest of the lattice, averaged over all sites. 

The local entropy becomes $s_1 =\textrm{Tr}_A ( \rho_A \ln \rho_A ) = -\sum_i p_i \ln(p_i)$. Physically, $s_1$ quantifies the uncertainty in the state of the single site. At finite temperature, $s_1$ is not a measure of pure quantum entanglement as it is contaminated by thermal contributions~\cite{Cardy:2017, Vedral:T2004, Anders:2007}.

Fig.~\ref{figS1} shows $s_1(\delta)$ as a function of doping for different values of $U$ at $T<(T_c^d)_{\rm max}$ both in the normal and superconducting states (open circles and and shaded squares, respectively). The analysis of $s_1$ across the {\it normal state} phase diagram of the two-dimensional Hubbard model has been performed in Refs.~\cite{Caitlin:PRL2019, Caitlin:PRXQ2020}. Upon doping a Mott insulator, $s_1(\delta)_{\rm NS}$ increases and then decreases with further doping, reflecting the competition between Mott localisation and Fermi statistics~\cite{Caitlin:PRXQ2020}. 
Up until now the behavior of $s_1$ in the {\it superconducting state} was unknown. 

Examining Fig.~\ref{figS1}, the superconducting and normal state local entropies differ, hence $s_1$ does detect the superconducting state. $(s_1)_{\rm SC}$ takes values larger or smaller than that in the normal state. The difference in $s_1$ between the superconducting and normal states (blue line in the insets of Fig.~\ref{figS1}) can be positive or negative. It is overall small (less than 1\%) and decreases with increasing $U$, but is crucially nonzero. 

It may seem surprising~\cite{amicoRMP2008, Gu:2004} that $s_1$, which depends on local quantities $D$ (the double occupancy) and $n$ (the occupancy) alone, is sensitive to the superconducting order. For example, it was suggested that the reduced density matrix of at least two sites is necessary to detect superconducting correlations~\cite{amicoRMP2008, Gu:2004, dengPRB2006}.
Note that for a given $n$, $s_1$ is a function of $D$ alone, and $DU$ is the potential energy of the Hubbard model. 
It has been shown in Refs.~\cite{maierENERGY, carbone2006, millisENERGY, LorenzoSC} that in the superconducting state the potential energy is lowered at large doping and increased at small doping, thus $\Delta E_{\rm pot}$ changes sign (dotted line in the insets of Fig.~\ref{figS1}). This change in potential energy is overcome by a kinetic energy change. Upon condensation, kinetic energy decreases at small dopings and may increase at large doping. Overall, this gives rise to a kinetic-energy driven superconductivity at small doping which progressively extends to larger doping with increasing $U$~\cite{LorenzoSC}. 
This observation explains the puzzling behavior of $s_1(\delta)_{\rm SC}$, which takes values larger or smaller than that in the normal state. The behavior of $s_1(\delta)$ in the superconducting state mirrors the behavior of the potential energy in the superconducting state, and in particular the change in $s_1$ upon condensation reflects the sign change of $\Delta E_{\rm pot}$ upon condensation - indeed the sign change occurs at the same doping~\cite{LorenzoSC}. 
Therefore, $s_1(\delta)$ not only detects the superconducting state, but also reflects the source of the condensation energy.

\subsection{Total mutual information and thermodynamic entropy}
Since $s_1$ contains thermal contributions, it is useful to also consider the mutual information between a single site and the rest of the lattice, which measures the total quantum and classical correlations shared between the site and its environment~\cite{watrous2018, groisman2005, wolfPRL2008, vidal:PRA2002}. For the site $i=1$, it is defined as $I(1:\left\{ > 1\right\}) = s_1 +s_{\left\{ > 1\right\}} - s_{\left\{ > 0\right\}}$, where $\left\{ > i \right\}$ is the set of sites with indices greater than $i$, and $\left\{ > 0 \right\}$ means the entire lattice. It is zero if the site and its environment are uncorrelated, and nonzero if they are correlated. 
Physically, $I(1:\left\{ > 1\right\}) \neq 0$ means that the entire lattice contains more information than the sum of its subsystems --here the single site $i=1$ and its environment. 
Simplifications and comparison with experiments with ultracold atoms in optical lattices~\cite{Cocchi:PRX2017} are possible by considering the {\it total} mutual information, which has been formally introduced in our Ref.~\cite{Caitlin:PRL2019} (see also Refs.~\cite{CaitlinSb, Caitlin:PRXQ2020}). 
It is defined as the mutual information between a site and the rest of the lattice, averaged over all the sites, $\overline{I}_1 = \frac{1}{N}\sum_{i=1}^N I(i: \left\{ >i \right\}) = \frac{1}{N}\sum_{i=1}^N [ s_1(i) + s_{\left\{ > i \right\}} -s_{\left\{ > i-1\right\}} ]$, where $N$ is the number of sites. 
This definition avoids double counting of the correlations between lattice sites~\cite{Caitlin:PRL2019}. It is easy to show~\cite{CaitlinSb, Caitlin:PRXQ2020} that most terms in the sum cancel, leaving $\overline{I}_1  = [\sum_{i=1}^N s_1(i)]/N -s$, where $s=s_{\left\{ > 0 \right\}}/N$ is the thermodynamic entropy per site. For translationally invariant systems, $\overline{I}_1$ reduces to the difference between the local and thermodynamic entropies, $\overline{I}_1 = s_1 -s$~\cite{Caitlin:PRL2019, Caitlin:PRXQ2020}.

\begin{figure}[t!]
\centering{
\includegraphics[width=0.98\linewidth]{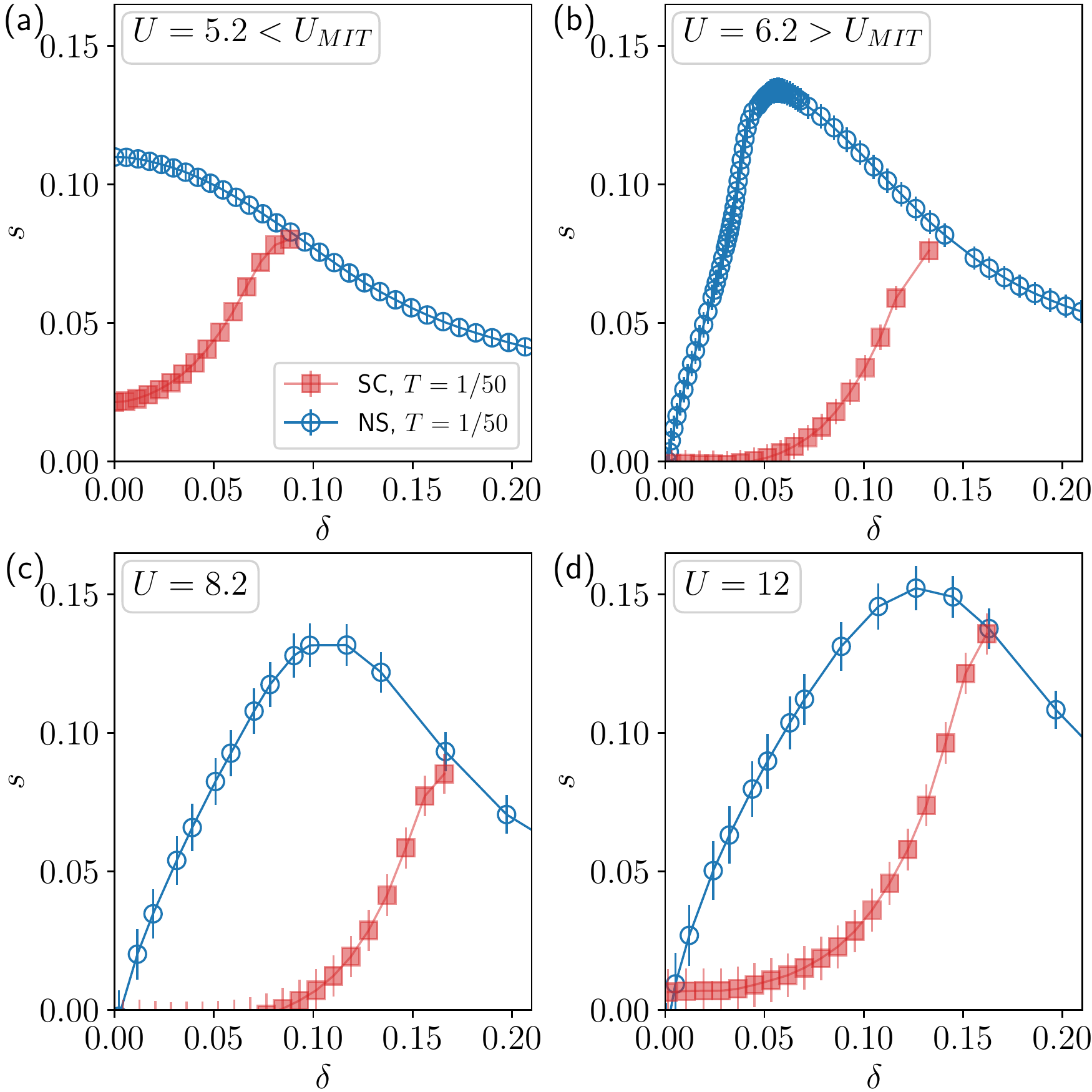}
}
\caption{Thermodynamic entropy $s$ as a function of doping, for (a) $U=5.2<U_{\rm MIT}$, (b) $U=6.2>U_{\rm MIT}$, (c) $U=8.2$, (d) $U=12$. Data in all panels are at $T=1/50$ for the normal and superconducting states (open circles and shaded squares respectively). Normal state data for $U=6.2$ are taken from Ref.~\cite{Caitlin:PRXQ2020}.
}
\label{figS}
\end{figure}

Thus, first let us consider the thermodynamic entropy $s$ as a function of doping as shown in Fig.~\ref{figS} for different values of $U$ at $T=1/50 <(T_c^d)_{\rm max}$ for both normal and superconducting states. It was obtained using the method of Ref.~\cite{CaitlinSb}, which exploits the Gibbs-Duhem relation (see Materials and Methods). The behavior of the thermodynamic entropy $s$ in the {\it normal} state has been studied in Ref.~\cite{Caitlin:PRXQ2020}. At high temperature, the results of Ref.~\cite{Caitlin:PRXQ2020} agree with experimental results with ultracold atoms~\cite{Cocchi:PRX2017}. At low temperature, they are compatible with the entropy landscape of cuprates~\cite{LoramJPCS2001, Tallon:2004}. The behavior of $s$ in the {\it superconducting} state up until now had not been reported within the two-dimensional Hubbard model with CDMFT. We expect our result to be a lower bound to the total entropy of the real system (see Materials and Methods). Here we compare $s$ in the superconducting and normal states. 

First, Fig.~\ref{figS} shows that the thermodynamic entropy is strongly suppressed in the superconducting state. This finding complements the suppression of the large scattering rate close to the antinode found in the normal state~\cite{LorenzoSC}. 
Physically, this is consistent with Cooper pairs locked into spin singlets propagating {\it coherently} in the lattice. 
Second, $s(\delta)_{\rm SC}$ monotonically decreases with decreasing doping for all values of $U$ considered here. Entropy becomes larger close to the high doping end of the superconducting dome where it must eventually recover its normal-state value. The low value of the entropy near half-filling in the doped Mott insulator regime $U>U_{\rm MIT}$ suggests that in this underdoped regime, singlets are more strongly bound by $J=4t^2/U$ than at larger doping. 
Third, the maximum in the entropy as a function of doping in the underlying normal state is completely obliterated by superconductivity. Hence, when considering stable states only, the maximum in entropy versus $\delta$ defines a vertical line in the $T-\delta$ phase diagram (gray crosses in Fig.~\ref{fig1}), which when it reaches $T_c^d$ then follows the high-doping side of the superconducting dome. 

Overall the behavior of $s$ in the superconducting and normal states as a function of doping and temperature provides a framework to interpret the experimental data on the entropy landscape of cuprates. For $U> U_{\rm MIT}$, qualitative features consistent with experiments in hole-doped cuprates~\cite{LoramJPCS2001, Tallon:2004} include: (a) a peak in $s(\delta)_{\rm NS}$ bounds the pseudogap region and is temperature independent (see Fig.~\ref{fig1}), and (b) a low value of $s(\delta)_{\rm SC}$ when superconductivity emerges from the underlying normal state pseudogap, followed by a rapid increase close to the high-doping end of the superconducting dome.

Entropy is a key quantity from which other thermodynamic properties can be derived. Examples include the specific heat $c_V$ and the thermopower. Under an intense magnetic field, experiments report a peak in $c_V$ at the doping where the pseudogap ends~\cite{Michon:Cv2018}. This is taken as a thermodynamic signature of a quantum critical point, but this hypothesis is complicated by the absence of a diverging correlation length associated to broken symmetry states. CDMFT study~\cite{Giovanni:PRBcv} solves this paradox by pinpointing the source of the specific heat anomaly as arising from the pseudogap to correlated metal critical point concealed by the superconducting state, without invoking broken symmetry states associated to the pseudogap. 
By the Kelvin formula~\cite{Shastry:2009}, when $s(\delta)$ is maximum, the thermopower changes sign, possibly suggesting a change of sign in the carriers at finite doping, as found in experiments~\cite{Honma:2008, ShilaPRB2010, CollignonPRB2021}. 

The findings of this work on $s$ add a contribution to the framework emerging from CDMFT studies, which suggests an explanation of the pseudogap phenomenology of hole-doped cuprates based on Mott physics and short range correlations.

Hole doped cuprates can be modelled by $U> U_{\rm MIT}$. On the other hand, it has been proposed~\cite{senechalPRL2004} that electron-doped cuprates can be modelled by a value of $U$ smaller than $U_{\rm MIT}$. In this regime, the entropy behavior of Fig.~\ref{figS}(a) is more relevant, in contrast with that emerging from a doped Mott insulator: there is no peak at finite doping in the $s(\delta)_{\rm NS}$ and there is a larger value of $s_{\rm SC}$. Therefore comparing the entropy landscape of electron vs hole-doped cuprates could also further support this view.

\begin{figure}[t!]
\centering{
\includegraphics[width=0.98\linewidth]{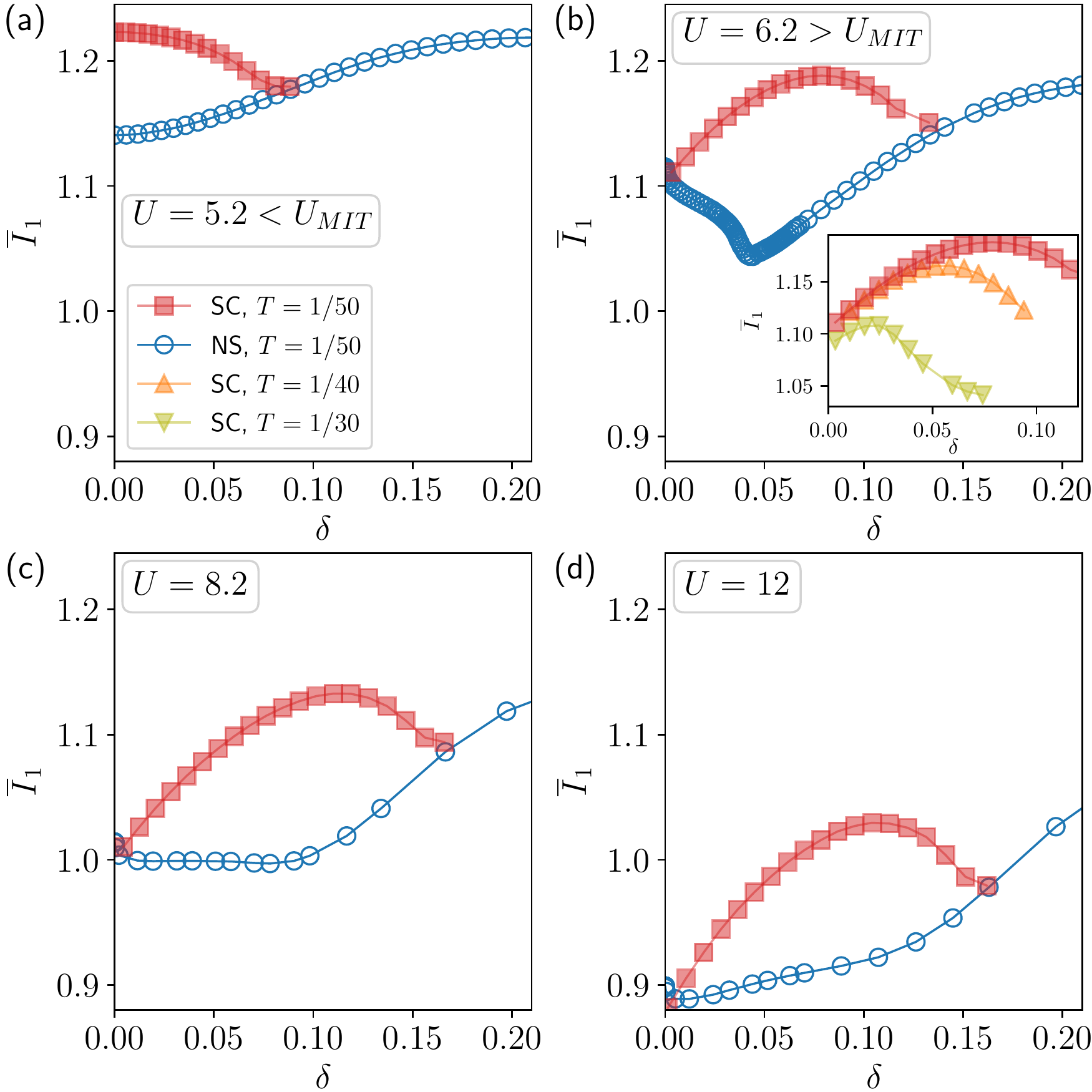}
}
\caption{Total mutual information $\overline{I}_1$ as a function of doping, for (a) $U=5.2<U_{\rm MIT}$, (b) $U=6.2>U_{\rm MIT}$, (c) $U=8.2$, (d) $U=12$. Data in all main panels are at $T=1/50$ for the normal and superconducting states (open circles and shaded squares respectively). Inset of panel (b) shows $(\overline{I}_1(\delta))_{\rm SC}$ at $U=6.2$ for $T=1/50$, $T=1/40$, and $T=1/30$ (squares, up triangles, and down triangles respectively). Normal state data for $U=6.2$ are reproduced from Ref.~\cite{Caitlin:PRXQ2020}.
}
\label{figI1}
\end{figure}

Next we turn to the total mutual information $\overline{I}_1$ versus $\delta$, shown in Fig.~\ref{figI1}. 
First, similarly to $s_1$, $\overline{I}_1$ does detect the superconducting state. However, upon condensation the changes in $\overline{I}_1$ are larger than those in $s_1$, and hence they could be detected more easily in experiments with ultracold atoms in optical lattices. 

Second, and contrary to $s_1$, $\overline{I}_1$ in the superconducting state is larger than $\overline{I}_1$ in the normal state for all doping levels and all values of $U$. This implies that the system in the superconducting state has more classical and quantum correlations than in the normal state.  
Since $\overline{I}_1$ and $s_1$ detect superconductivity in different ways, this could also suggest the possibility that different correlations encoded in $\overline{I}_1$ and not in $s_1$ contribute to superconductivity. Note that $\overline{I}_1$ and $s_1$ detect the pseudogap to correlated metal transition in the underlying normal state in a similar way: they both show an inflection vs chemical potential~\cite{Caitlin:PRXQ2020}. Distinguishing pure quantum from classical correlations could offer further insight into the nature of the correlations in the superconducting state. However, this is not possible with the entanglement-related quantities $\overline{I}_1$ and $s_1$, and remains an unsolved problem in the information theory description of finite temperature phase transitions~\cite{zhengBOOK, Wald:JSM2020}. 

Third, $(\overline{I}_1)_{\rm SC}$ is sensitive to the critical value $U_{\rm MIT}$. 
For $U<U_{\rm MIT}$, $(\overline{I}_1(\delta))_{\rm SC}$ decreases with increasing $\delta$. In contrast, for $U>U_{\rm MIT}$, $(\overline{I}_1(\delta))_{\rm SC}$ first increases and then decreases with increasing $\delta$, and thus has a broad maximum at finite doping. This implies that the superconductivity emerging from a doped correlated metal ($U<U_{\rm MIT}$) or a doped Mott insulator ($U>U_{\rm MIT}$) leaves a distinctive feature in $(\overline{I}_1(\delta))_{\rm SC}$. 

In particular, this maximum in $(\overline{I}_1(\delta))_{\rm SC}$ replaces the minimum that exists at finite doping in $(\overline{I}_1(\delta))_{\rm NS}$, which is a result of the competition between spin and charge correlations~\cite{Caitlin:PRXQ2020}. 
The spin correlations are caused by $J$ and win at low $\delta$, whereas the charge correlations win at large $\delta$. 
Mathematically, the maximum in $(\overline{I}_1(\delta))_{\rm SC}$ can be understood by the fact that upon doping the Mott state, $(\overline{I}_1)_{\rm SC}$ is governed by $(s_1)_{\rm SC}$ since $(s)_{\rm SC}$ is essentially zero, and hence $(\overline{I}_1)_{\rm SC}$ increases with $\delta$ as $s_1(\delta)$ does. Then, upon approaching the high-doping end of the superconducting dome, $(s)_{\rm SC}$ grows with increasing $\delta$, and hence $(\overline{I}_1)_{\rm SC}$ decreases with increasing $\delta$. 
Physically, the broad peak arises because of the competition between quantum aspects encoded in $s_1$ that win at low $\delta$, and classical aspects encoded in $s$ that win at larger $\delta$. This is further supported by the behavior of $(\overline{I}_1)_{\rm SC}$ for different temperatures, as shown in the inset of Fig.~\ref{figI1}b for $U=6.2$. The lower the temperature, the broader the range in doping where $(\overline{I}_1(\delta))_{\rm SC}$ increases with increasing $\delta$.

\section{Conclusions}
In summary, we use information theoretic measures to study quantum and classical correlations in the superconducting state of the finite-temperature two-dimensional Hubbard model with CDMFT, by tuning temperature, doping and interaction strength. 
Thermodynamic entropy $s$ at finite temperature is very strongly suppressed near half-filling in the superconducting state. By calculating the local entropy and total mutual information between a site and its environment, we show that both $s_1$ and $\overline{I}_1$ detect the superconducting state. The changes in $s_1(\delta)$ upon condensation follow the changes in potential energy and thus can be positive or negative. In sharp contrast, $\overline{I}_1$ in the superconducting state is larger than in the normal state at all doping levels and for all $U$, revealing increased correlations in the superconducting state, and shows a broad peak versus doping in the doped Mott insulator case, as a result of competing quantum and classical effects. 

On the theory level, these findings pave the way for a deeper understanding of superconductivity emerging from a doped Mott insulator, using information-theory tools which complement thermodynamic tools~\cite{zhengBOOK}. On the experimental side, the theoretical framework of our work can be applied to ultracold atom realisations of the Hubbard model. Our predictions could be tested soon, when superconducting correlations can be detected in these systems. Further, our findings on the thermodynamic entropy immediately offer a microscopic model to better our understanding of the entropy landscape of cuprates~\cite{LoramJPCS2001, Tallon:2004}.


\section*{Materials and methods}
\subsection*{CDMFT simulations} We solve the cluster quantum impurity model associated to the CDMFT method using the hybridisation expansion continuous time quantum Monte Carlo method~\cite{millisRMP, patrickSkipList}. We use 72 to 96 processors, with about $2-4 \times 10^7$ Monte Carlo steps per processor. The CDMFT self-consistency is iterated until convergence, which is typically reached within 50 iterations, although hundreds of iterations are necessary close to the superconducting phase boundaries or to the pseudogap to correlated metal boundaries and its associated crossovers. Once convergence is reached, we take the average of between 30 to several hundred CDMFT iterations. Our goal is that the standard deviation of local observables such as the occupancy $n$ and the double occupancy $D$ is of the order $10^{-5}$.  

SI Appendix Fig.~S2 shows the resulting average sign of the Monte Carlo simulations as a function of doping for the set of parameters used in Figs.~\ref{figS1}, \ref{figS}, \ref{figI1}. In particular, the temperature $T=1/50$ is the lowest temperature attained in this work. It is chosen since it is well below $(T_c^d)_{\rm max}$ for all values of $U$ considered here, minimizing the effect of vortex fluctuations on entropy, and because the sign problem worsens with decreasing temperature.

\subsection*{Construction of thermodynamic entropy} To obtain the thermodynamic entropy $s$, we use the protocol described in Ref.~\cite{CaitlinSb}. It is based on the Gibbs-Duhem relation $sdT -adP +nd\mu=0$, where $a$ is the area per site and $P$ is the pressure. This protocol only requires the knowledge of the occupation $n(\mu)$ from the empty band to the half filled band, from which $P$ and then $s$ are extracted, described as follows and illustrated in SI Appendix Fig.~S3. 

At fixed $T$ and $U$, the Gibbs-Duhem relation reduces to $nd\mu=adP$. By integration over $\mu$, we obtain the pressure $P(\mu)=1/a \int_{-\infty}^{\mu} n(\mu')_T d\mu'$. To perform the numerical integration, we use the composite trapezoidal method and a lower limit of integration such that $n(\mu_{\rm min}) \approx 0.002$. Due to the subtle variations in $n(\mu)_T$, it is crucial to obtain $n$ on small steps of $\mu$, which are from 0.2 down to 0.025. Superconductivity is found in a limited subset of $\mu$ (or doping, see Fig.~\ref{fig1}). Therefore, in order to get $P$ in the superconducting state, this integral is split into two parts: $P(\mu)_{\rm SC} = \int_{-\infty}^{\mu_{\rm SC}} n(\mu')_T d\mu' + \int_{\mu_{\rm SC}}^{\mu} n(\mu')_T d\mu'$, where $\mu_{\rm SC}$ is the lowest value of $\mu$ in the superconducting state. 

Finally, at fixed $\mu$ the Gibbs-Duhem relation gives the thermodynamic entropy per site $s(\mu)_T = a \left( dP/dT \right)_\mu$. To perform the numerical derivative, we take finite differences between two temperatures. We estimate $s$ at $T=1/50$ shown in Fig.~\ref{figS} by taking finite differences between pressure curves $P(\mu)$ at $T=1/40$ and $T=1/50$. 

\subsection*{Data availability} All relevant data are included in the article and SI Appendix. The datasets and the codes used for data analysis are available at the Open Science Framework repository \url{https://osf.io/a8b4y/}~\cite{dataSCENT}.


\section*{Acknowledgments}
This work has been supported by the Natural Sciences and Engineering Research Council of Canada (NSERC) under grants RGPIN-2019-05312, the Canada First Research Excellence Fund and by the Research Chair in the Theory of Quantum Materials. PS work was supported by the U.S. Department of Energy, Office of Science, Basic Energy Sciences as a part of the Computational Materials Science Program. Simulations were performed on computers provided by the Canadian Foundation for Innovation, the Minist\`ere de l'\'Education des Loisirs et du Sport (Qu\'ebec), Calcul Qu\'ebec, and Compute Canada.


\begin{thebibliography}{61}%
\makeatletter
\providecommand \@ifxundefined [1]{%
 \@ifx{#1\undefined}
}%
\providecommand \@ifnum [1]{%
 \ifnum #1\expandafter \@firstoftwo
 \else \expandafter \@secondoftwo
 \fi
}%
\providecommand \@ifx [1]{%
 \ifx #1\expandafter \@firstoftwo
 \else \expandafter \@secondoftwo
 \fi
}%
\providecommand \natexlab [1]{#1}%
\providecommand \enquote  [1]{``#1''}%
\providecommand \bibnamefont  [1]{#1}%
\providecommand \bibfnamefont [1]{#1}%
\providecommand \citenamefont [1]{#1}%
\providecommand \href@noop [0]{\@secondoftwo}%
\providecommand \href [0]{\begingroup \@sanitize@url \@href}%
\providecommand \@href[1]{\@@startlink{#1}\@@href}%
\providecommand \@@href[1]{\endgroup#1\@@endlink}%
\providecommand \@sanitize@url [0]{\catcode `\\12\catcode `\$12\catcode
  `\&12\catcode `\#12\catcode `\^12\catcode `\_12\catcode `\%12\relax}%
\providecommand \@@startlink[1]{}%
\providecommand \@@endlink[0]{}%
\providecommand \url  [0]{\begingroup\@sanitize@url \@url }%
\providecommand \@url [1]{\endgroup\@href {#1}{\urlprefix }}%
\providecommand \urlprefix  [0]{URL }%
\providecommand \Eprint [0]{\href }%
\providecommand \doibase [0]{http://dx.doi.org/}%
\providecommand \selectlanguage [0]{\@gobble}%
\providecommand \bibinfo  [0]{\@secondoftwo}%
\providecommand \bibfield  [0]{\@secondoftwo}%
\providecommand \translation [1]{[#1]}%
\providecommand \BibitemOpen [0]{}%
\providecommand \bibitemStop [0]{}%
\providecommand \bibitemNoStop [0]{.\EOS\space}%
\providecommand \EOS [0]{\spacefactor3000\relax}%
\providecommand \BibitemShut  [1]{\csname bibitem#1\endcsname}%
\let\auto@bib@innerbib\@empty
\bibitem [{\citenamefont {Zeng}\ \emph {et~al.}(2019)\citenamefont {Zeng},
  \citenamefont {Chen}, \citenamefont {Zhou},\ and\ \citenamefont
  {Wen}}]{zhengBOOK}%
  \BibitemOpen
  \bibfield  {author} {\bibinfo {author} {\bibfnamefont {B.}~\bibnamefont
  {Zeng}}, \bibinfo {author} {\bibfnamefont {X.}~\bibnamefont {Chen}}, \bibinfo
  {author} {\bibfnamefont {D.-L.}\ \bibnamefont {Zhou}}, \ and\ \bibinfo
  {author} {\bibfnamefont {X.-G.}\ \bibnamefont {Wen}},\ }\href {\doibase
  10.1007/978-1-4939-9084-9} {\emph {\bibinfo {title} {{Quantum Information
  Meets Quantum Matter}}}}\ (\bibinfo  {publisher} {Springer-Verlag},\ \bibinfo
  {address} {New York},\ \bibinfo {year} {2019})\BibitemShut {NoStop}%
\bibitem [{\citenamefont {Anderson}(1987)}]{Anderson:1987}%
  \BibitemOpen
  \bibfield  {author} {\bibinfo {author} {\bibfnamefont {P.~W.}\ \bibnamefont
  {Anderson}},\ }\bibfield  {title} {\enquote {\bibinfo {title} {{The
  resonating valence bond state in La$_2$CuO$_4$ and superconductivity}},}\
  }\href {\doibase 10.1126/science.235.4793.1196} {\bibfield  {journal}
  {\bibinfo  {journal} {Science}\ }\textbf {\bibinfo {volume} {235}},\ \bibinfo
  {pages} {1196--1198} (\bibinfo {year} {1987})}\BibitemShut {NoStop}%
\bibitem [{\citenamefont {Lee}\ \emph {et~al.}(2006)\citenamefont {Lee},
  \citenamefont {Nagaosa},\ and\ \citenamefont {Wen}}]{lee}%
  \BibitemOpen
  \bibfield  {author} {\bibinfo {author} {\bibfnamefont {Patrick~A.}\
  \bibnamefont {Lee}}, \bibinfo {author} {\bibfnamefont {Naoto}\ \bibnamefont
  {Nagaosa}}, \ and\ \bibinfo {author} {\bibfnamefont {Xiao-Gang}\ \bibnamefont
  {Wen}},\ }\bibfield  {title} {\enquote {\bibinfo {title} {Doping a mott
  insulator: Physics of high-temperature superconductivity},}\ }\href {\doibase
  10.1103/RevModPhys.78.17} {\bibfield  {journal} {\bibinfo  {journal} {Rev.
  Mod. Phys.}\ }\textbf {\bibinfo {volume} {78}},\ \bibinfo {eid} {17}
  (\bibinfo {year} {2006})}\BibitemShut {NoStop}%
\bibitem [{\citenamefont {Tremblay}\ \emph {et~al.}(2006)\citenamefont
  {Tremblay}, \citenamefont {Kyung},\ and\ \citenamefont
  {S\'{e}n\'{e}chal}}]{tremblayR}%
  \BibitemOpen
  \bibfield  {author} {\bibinfo {author} {\bibfnamefont {A.-M.~S.}\
  \bibnamefont {Tremblay}}, \bibinfo {author} {\bibfnamefont {B.}~\bibnamefont
  {Kyung}}, \ and\ \bibinfo {author} {\bibfnamefont {D.}~\bibnamefont
  {S\'{e}n\'{e}chal}},\ }\bibfield  {title} {\enquote {\bibinfo {title}
  {{Pseudogap and high-temperature superconductivity from weak to strong
  coupling. Towards a quantitative theory}},}\ }\href {\doibase
  10.1063/1.2199446} {\bibfield  {journal} {\bibinfo  {journal} {Low Temp.
  Phys.}\ }\textbf {\bibinfo {volume} {32}},\ \bibinfo {pages} {424} (\bibinfo
  {year} {2006})}\BibitemShut {NoStop}%
\bibitem [{\citenamefont {Alloul}(2014)}]{Alloul2013}%
  \BibitemOpen
  \bibfield  {author} {\bibinfo {author} {\bibfnamefont {Henri}\ \bibnamefont
  {Alloul}},\ }\bibfield  {title} {\enquote {\bibinfo {title} {What is the
  simplest model that captures the basic experimental facts of the physics of
  underdoped cuprates?}}\ }\href {\doibase 10.1016/j.crhy.2014.02.007}
  {\bibfield  {journal} {\bibinfo  {journal} {Comptes Rendus Physique}\
  }\textbf {\bibinfo {volume} {15}},\ \bibinfo {pages} {519 -- 524} (\bibinfo
  {year} {2014})}\BibitemShut {NoStop}%
\bibitem [{\citenamefont {Jaksch}\ and\ \citenamefont {Zoller}(2005)}]{jzHM}%
  \BibitemOpen
  \bibfield  {author} {\bibinfo {author} {\bibfnamefont {D.}~\bibnamefont
  {Jaksch}}\ and\ \bibinfo {author} {\bibfnamefont {P.}~\bibnamefont
  {Zoller}},\ }\bibfield  {title} {\enquote {\bibinfo {title} {The cold atom
  hubbard toolbox},}\ }\href {\doibase
  https://doi.org/10.1016/j.aop.2004.09.010} {\bibfield  {journal} {\bibinfo
  {journal} {Annals of Physics}\ }\textbf {\bibinfo {volume} {315}},\ \bibinfo
  {pages} {52--79} (\bibinfo {year} {2005})},\ \bibinfo {note} {special
  Issue}\BibitemShut {NoStop}%
\bibitem [{\citenamefont {Esslinger}(2010)}]{Esslinger:2010}%
  \BibitemOpen
  \bibfield  {author} {\bibinfo {author} {\bibfnamefont {Tilman}\ \bibnamefont
  {Esslinger}},\ }\bibfield  {title} {\enquote {\bibinfo {title}
  {{Fermi-Hubbard Physics with Atoms in an Optical Lattice}},}\ }\href
  {\doibase 10.1146/annurev-conmatphys-070909-104059} {\bibfield  {journal}
  {\bibinfo  {journal} {Annual Review of Condensed Matter Physics}\ }\textbf
  {\bibinfo {volume} {1}},\ \bibinfo {pages} {129--152} (\bibinfo {year}
  {2010})}\BibitemShut {NoStop}%
\bibitem [{\citenamefont {Gross}\ and\ \citenamefont
  {Bloch}(2017)}]{GrossScience2017}%
  \BibitemOpen
  \bibfield  {author} {\bibinfo {author} {\bibfnamefont {Christian}\
  \bibnamefont {Gross}}\ and\ \bibinfo {author} {\bibfnamefont {Immanuel}\
  \bibnamefont {Bloch}},\ }\bibfield  {title} {\enquote {\bibinfo {title}
  {Quantum simulations with ultracold atoms in optical lattices},}\ }\href
  {\doibase 10.1126/science.aal3837} {\bibfield  {journal} {\bibinfo  {journal}
  {Science}\ }\textbf {\bibinfo {volume} {357}},\ \bibinfo {pages} {995--1001}
  (\bibinfo {year} {2017})}\BibitemShut {NoStop}%
\bibitem [{\citenamefont {Amico}\ \emph {et~al.}(2008)\citenamefont {Amico},
  \citenamefont {Fazio}, \citenamefont {Osterloh},\ and\ \citenamefont
  {Vedral}}]{amicoRMP2008}%
  \BibitemOpen
  \bibfield  {author} {\bibinfo {author} {\bibfnamefont {Luigi}\ \bibnamefont
  {Amico}}, \bibinfo {author} {\bibfnamefont {Rosario}\ \bibnamefont {Fazio}},
  \bibinfo {author} {\bibfnamefont {Andreas}\ \bibnamefont {Osterloh}}, \ and\
  \bibinfo {author} {\bibfnamefont {Vlatko}\ \bibnamefont {Vedral}},\
  }\bibfield  {title} {\enquote {\bibinfo {title} {Entanglement in many-body
  systems},}\ }\href {\doibase 10.1103/RevModPhys.80.517} {\bibfield  {journal}
  {\bibinfo  {journal} {Rev. Mod. Phys.}\ }\textbf {\bibinfo {volume} {80}},\
  \bibinfo {pages} {517--576} (\bibinfo {year} {2008})}\BibitemShut {NoStop}%
\bibitem [{\citenamefont {Laflorencie}(2016)}]{Laflorencie:PhysRep2016}%
  \BibitemOpen
  \bibfield  {author} {\bibinfo {author} {\bibfnamefont {Nicolas}\ \bibnamefont
  {Laflorencie}},\ }\bibfield  {title} {\enquote {\bibinfo {title} {Quantum
  entanglement in condensed matter systems},}\ }\href {\doibase
  https://doi.org/10.1016/j.physrep.2016.06.008} {\bibfield  {journal}
  {\bibinfo  {journal} {Physics Reports}\ }\textbf {\bibinfo {volume} {646}},\
  \bibinfo {pages} {1--59} (\bibinfo {year} {2016})}\BibitemShut {NoStop}%
\bibitem [{\citenamefont {Islam}\ \emph {et~al.}(2015)\citenamefont {Islam},
  \citenamefont {Ma}, \citenamefont {Preiss}, \citenamefont {Tai},
  \citenamefont {Lukin}, \citenamefont {Rispoli},\ and\ \citenamefont
  {Greiner}}]{greinerNat2015}%
  \BibitemOpen
  \bibfield  {author} {\bibinfo {author} {\bibfnamefont {Rajibul}\ \bibnamefont
  {Islam}}, \bibinfo {author} {\bibfnamefont {Ruichao}\ \bibnamefont {Ma}},
  \bibinfo {author} {\bibfnamefont {Philipp~M}\ \bibnamefont {Preiss}},
  \bibinfo {author} {\bibfnamefont {M~Eric}\ \bibnamefont {Tai}}, \bibinfo
  {author} {\bibfnamefont {Alexander}\ \bibnamefont {Lukin}}, \bibinfo {author}
  {\bibfnamefont {Matthew}\ \bibnamefont {Rispoli}}, \ and\ \bibinfo {author}
  {\bibfnamefont {Markus}\ \bibnamefont {Greiner}},\ }\bibfield  {title}
  {\enquote {\bibinfo {title} {Measuring entanglement entropy in a quantum
  many-body system},}\ }\href {\doibase 10.1038/nature15750} {\bibfield
  {journal} {\bibinfo  {journal} {Nature}\ }\textbf {\bibinfo {volume} {528}},\
  \bibinfo {pages} {77} (\bibinfo {year} {2015})}\BibitemShut {NoStop}%
\bibitem [{\citenamefont {Kaufman}\ \emph {et~al.}(2016)\citenamefont
  {Kaufman}, \citenamefont {Tai}, \citenamefont {Lukin}, \citenamefont
  {Rispoli}, \citenamefont {Schittko}, \citenamefont {Preiss},\ and\
  \citenamefont {Greiner}}]{Kaufman:Science2016}%
  \BibitemOpen
  \bibfield  {author} {\bibinfo {author} {\bibfnamefont {Adam~M.}\ \bibnamefont
  {Kaufman}}, \bibinfo {author} {\bibfnamefont {M.~Eric}\ \bibnamefont {Tai}},
  \bibinfo {author} {\bibfnamefont {Alexander}\ \bibnamefont {Lukin}}, \bibinfo
  {author} {\bibfnamefont {Matthew}\ \bibnamefont {Rispoli}}, \bibinfo {author}
  {\bibfnamefont {Robert}\ \bibnamefont {Schittko}}, \bibinfo {author}
  {\bibfnamefont {Philipp~M.}\ \bibnamefont {Preiss}}, \ and\ \bibinfo {author}
  {\bibfnamefont {Markus}\ \bibnamefont {Greiner}},\ }\bibfield  {title}
  {\enquote {\bibinfo {title} {Quantum thermalization through entanglement in
  an isolated many-body system},}\ }\href {\doibase 10.1126/science.aaf6725}
  {\bibfield  {journal} {\bibinfo  {journal} {Science}\ }\textbf {\bibinfo
  {volume} {353}},\ \bibinfo {pages} {794--800} (\bibinfo {year}
  {2016})}\BibitemShut {NoStop}%
\bibitem [{\citenamefont {Cocchi}\ \emph {et~al.}(2017)\citenamefont {Cocchi},
  \citenamefont {Miller}, \citenamefont {Drewes}, \citenamefont {Chan},
  \citenamefont {Pertot}, \citenamefont {Brennecke},\ and\ \citenamefont
  {K\"ohl}}]{Cocchi:PRX2017}%
  \BibitemOpen
  \bibfield  {author} {\bibinfo {author} {\bibfnamefont {E.}~\bibnamefont
  {Cocchi}}, \bibinfo {author} {\bibfnamefont {L.~A.}\ \bibnamefont {Miller}},
  \bibinfo {author} {\bibfnamefont {J.~H.}\ \bibnamefont {Drewes}}, \bibinfo
  {author} {\bibfnamefont {C.~F.}\ \bibnamefont {Chan}}, \bibinfo {author}
  {\bibfnamefont {D.}~\bibnamefont {Pertot}}, \bibinfo {author} {\bibfnamefont
  {F.}~\bibnamefont {Brennecke}}, \ and\ \bibinfo {author} {\bibfnamefont
  {M.}~\bibnamefont {K\"ohl}},\ }\bibfield  {title} {\enquote {\bibinfo {title}
  {{Measuring Entropy and Short-Range Correlations in the Two-Dimensional
  Hubbard Model}},}\ }\href {\doibase 10.1103/PhysRevX.7.031025} {\bibfield
  {journal} {\bibinfo  {journal} {Phys. Rev. X}\ }\textbf {\bibinfo {volume}
  {7}},\ \bibinfo {pages} {031025} (\bibinfo {year} {2017})}\BibitemShut
  {NoStop}%
\bibitem [{\citenamefont {Lukin}\ \emph {et~al.}(2019)\citenamefont {Lukin},
  \citenamefont {Rispoli}, \citenamefont {Schittko}, \citenamefont {Tai},
  \citenamefont {Kaufman}, \citenamefont {Choi}, \citenamefont {Khemani},
  \citenamefont {L{\'e}onard},\ and\ \citenamefont
  {Greiner}}]{Lukin:Science2019}%
  \BibitemOpen
  \bibfield  {author} {\bibinfo {author} {\bibfnamefont {Alexander}\
  \bibnamefont {Lukin}}, \bibinfo {author} {\bibfnamefont {Matthew}\
  \bibnamefont {Rispoli}}, \bibinfo {author} {\bibfnamefont {Robert}\
  \bibnamefont {Schittko}}, \bibinfo {author} {\bibfnamefont {M.~Eric}\
  \bibnamefont {Tai}}, \bibinfo {author} {\bibfnamefont {Adam~M.}\ \bibnamefont
  {Kaufman}}, \bibinfo {author} {\bibfnamefont {Soonwon}\ \bibnamefont {Choi}},
  \bibinfo {author} {\bibfnamefont {Vedika}\ \bibnamefont {Khemani}}, \bibinfo
  {author} {\bibfnamefont {Julian}\ \bibnamefont {L{\'e}onard}}, \ and\
  \bibinfo {author} {\bibfnamefont {Markus}\ \bibnamefont {Greiner}},\
  }\bibfield  {title} {\enquote {\bibinfo {title} {Probing entanglement in a
  many-body{\textendash}localized system},}\ }\href {\doibase
  10.1126/science.aau0818} {\bibfield  {journal} {\bibinfo  {journal}
  {Science}\ }\textbf {\bibinfo {volume} {364}},\ \bibinfo {pages} {256--260}
  (\bibinfo {year} {2019})}\BibitemShut {NoStop}%
\bibitem [{\citenamefont {Hofstetter}\ \emph {et~al.}(2002)\citenamefont
  {Hofstetter}, \citenamefont {Cirac}, \citenamefont {Zoller}, \citenamefont
  {Demler},\ and\ \citenamefont {Lukin}}]{Hofstetter:2002}%
  \BibitemOpen
  \bibfield  {author} {\bibinfo {author} {\bibfnamefont {W.}~\bibnamefont
  {Hofstetter}}, \bibinfo {author} {\bibfnamefont {J.~I.}\ \bibnamefont
  {Cirac}}, \bibinfo {author} {\bibfnamefont {P.}~\bibnamefont {Zoller}},
  \bibinfo {author} {\bibfnamefont {E.}~\bibnamefont {Demler}}, \ and\ \bibinfo
  {author} {\bibfnamefont {M.~D.}\ \bibnamefont {Lukin}},\ }\bibfield  {title}
  {\enquote {\bibinfo {title} {High-temperature superfluidity of fermionic
  atoms in optical lattices},}\ }\href {\doibase 10.1103/PhysRevLett.89.220407}
  {\bibfield  {journal} {\bibinfo  {journal} {Phys. Rev. Lett.}\ }\textbf
  {\bibinfo {volume} {89}},\ \bibinfo {pages} {220407} (\bibinfo {year}
  {2002})}\BibitemShut {NoStop}%
\bibitem [{\citenamefont {Jaynes}(1957)}]{Jaynes:1957}%
  \BibitemOpen
  \bibfield  {author} {\bibinfo {author} {\bibfnamefont {E.~T.}\ \bibnamefont
  {Jaynes}},\ }\bibfield  {title} {\enquote {\bibinfo {title} {{Information
  Theory and Statistical Mechanics}},}\ }\href {\doibase
  10.1103/PhysRev.106.620} {\bibfield  {journal} {\bibinfo  {journal} {Phys.
  Rev.}\ }\textbf {\bibinfo {volume} {106}},\ \bibinfo {pages} {620--630}
  (\bibinfo {year} {1957})}\BibitemShut {NoStop}%
\bibitem [{\citenamefont {Watrous}(2018)}]{watrous2018}%
  \BibitemOpen
  \bibfield  {author} {\bibinfo {author} {\bibfnamefont {John}\ \bibnamefont
  {Watrous}},\ }\href@noop {} {\emph {\bibinfo {title} {The Theory of Quantum
  Information}}}\ (\bibinfo  {publisher} {Cambridge University Press},\
  \bibinfo {year} {2018})\BibitemShut {NoStop}%
\bibitem [{\citenamefont {Walsh}\ \emph {et~al.}(2020)\citenamefont {Walsh},
  \citenamefont {S\'emon}, \citenamefont {Poulin}, \citenamefont {Sordi},\ and\
  \citenamefont {Tremblay}}]{Caitlin:PRXQ2020}%
  \BibitemOpen
  \bibfield  {author} {\bibinfo {author} {\bibfnamefont {C.}~\bibnamefont
  {Walsh}}, \bibinfo {author} {\bibfnamefont {P.}~\bibnamefont {S\'emon}},
  \bibinfo {author} {\bibfnamefont {D.}~\bibnamefont {Poulin}}, \bibinfo
  {author} {\bibfnamefont {G.}~\bibnamefont {Sordi}}, \ and\ \bibinfo {author}
  {\bibfnamefont {A.-M.~S.}\ \bibnamefont {Tremblay}},\ }\bibfield  {title}
  {\enquote {\bibinfo {title} {Entanglement and classical correlations at the
  doping-driven mott transition in the two-dimensional hubbard model},}\ }\href
  {\doibase 10.1103/PRXQuantum.1.020310} {\bibfield  {journal} {\bibinfo
  {journal} {PRX Quantum}\ }\textbf {\bibinfo {volume} {1}},\ \bibinfo {pages}
  {020310} (\bibinfo {year} {2020})}\BibitemShut {NoStop}%
\bibitem [{\citenamefont {Maier}\ \emph {et~al.}(2005)\citenamefont {Maier},
  \citenamefont {Jarrell}, \citenamefont {Pruschke},\ and\ \citenamefont
  {Hettler}}]{maier}%
  \BibitemOpen
  \bibfield  {author} {\bibinfo {author} {\bibfnamefont {Thomas}\ \bibnamefont
  {Maier}}, \bibinfo {author} {\bibfnamefont {Mark}\ \bibnamefont {Jarrell}},
  \bibinfo {author} {\bibfnamefont {Thomas}\ \bibnamefont {Pruschke}}, \ and\
  \bibinfo {author} {\bibfnamefont {Matthias~H.}\ \bibnamefont {Hettler}},\
  }\bibfield  {title} {\enquote {\bibinfo {title} {Quantum cluster theories},}\
  }\href {\doibase 10.1103/RevModPhys.77.1027} {\bibfield  {journal} {\bibinfo
  {journal} {Rev. Mod. Phys.}\ }\textbf {\bibinfo {volume} {77}},\ \bibinfo
  {pages} {1027--1080} (\bibinfo {year} {2005})}\BibitemShut {NoStop}%
\bibitem [{\citenamefont {Kotliar}\ \emph {et~al.}(2006)\citenamefont
  {Kotliar}, \citenamefont {Savrasov}, \citenamefont {Haule}, \citenamefont
  {Oudovenko}, \citenamefont {Parcollet},\ and\ \citenamefont
  {Marianetti}}]{kotliarRMP}%
  \BibitemOpen
  \bibfield  {author} {\bibinfo {author} {\bibfnamefont {G.}~\bibnamefont
  {Kotliar}}, \bibinfo {author} {\bibfnamefont {S.~Y.}\ \bibnamefont
  {Savrasov}}, \bibinfo {author} {\bibfnamefont {K.}~\bibnamefont {Haule}},
  \bibinfo {author} {\bibfnamefont {V.~S.}\ \bibnamefont {Oudovenko}}, \bibinfo
  {author} {\bibfnamefont {O.}~\bibnamefont {Parcollet}}, \ and\ \bibinfo
  {author} {\bibfnamefont {C.~A.}\ \bibnamefont {Marianetti}},\ }\bibfield
  {title} {\enquote {\bibinfo {title} {{Electronic structure calculations with
  dynamical mean-field theory}},}\ }\href {\doibase 10.1103/RevModPhys.78.865}
  {\bibfield  {journal} {\bibinfo  {journal} {Rev. Mod. Phys.}\ }\textbf
  {\bibinfo {volume} {78}},\ \bibinfo {eid} {865} (\bibinfo {year}
  {2006})}\BibitemShut {NoStop}%
\bibitem [{\citenamefont {Georges}\ \emph {et~al.}(1996)\citenamefont
  {Georges}, \citenamefont {Kotliar}, \citenamefont {Krauth},\ and\
  \citenamefont {Rozenberg}}]{rmp}%
  \BibitemOpen
  \bibfield  {author} {\bibinfo {author} {\bibfnamefont {Antoine}\ \bibnamefont
  {Georges}}, \bibinfo {author} {\bibfnamefont {Gabriel}\ \bibnamefont
  {Kotliar}}, \bibinfo {author} {\bibfnamefont {Werner}\ \bibnamefont
  {Krauth}}, \ and\ \bibinfo {author} {\bibfnamefont {Marcelo~J.}\ \bibnamefont
  {Rozenberg}},\ }\bibfield  {title} {\enquote {\bibinfo {title} {{Dynamical
  mean-field theory of strongly correlated fermion systems and the limit of
  infinite dimensions}},}\ }\href {\doibase 10.1103/RevModPhys.68.13}
  {\bibfield  {journal} {\bibinfo  {journal} {Rev. Mod. Phys.}\ }\textbf
  {\bibinfo {volume} {68}},\ \bibinfo {pages} {13} (\bibinfo {year}
  {1996})}\BibitemShut {NoStop}%
\bibitem [{\citenamefont {Gull}\ \emph {et~al.}(2011)\citenamefont {Gull},
  \citenamefont {Millis}, \citenamefont {Lichtenstein}, \citenamefont
  {Rubtsov}, \citenamefont {Troyer},\ and\ \citenamefont {Werner}}]{millisRMP}%
  \BibitemOpen
  \bibfield  {author} {\bibinfo {author} {\bibfnamefont {Emanuel}\ \bibnamefont
  {Gull}}, \bibinfo {author} {\bibfnamefont {Andrew~J.}\ \bibnamefont
  {Millis}}, \bibinfo {author} {\bibfnamefont {Alexander~I.}\ \bibnamefont
  {Lichtenstein}}, \bibinfo {author} {\bibfnamefont {Alexey~N.}\ \bibnamefont
  {Rubtsov}}, \bibinfo {author} {\bibfnamefont {Matthias}\ \bibnamefont
  {Troyer}}, \ and\ \bibinfo {author} {\bibfnamefont {Philipp}\ \bibnamefont
  {Werner}},\ }\bibfield  {title} {\enquote {\bibinfo {title} {{Continuous-time
  Monte~Carlo methods for quantum impurity models}},}\ }\href {\doibase
  10.1103/RevModPhys.83.349} {\bibfield  {journal} {\bibinfo  {journal} {Rev.
  Mod. Phys.}\ }\textbf {\bibinfo {volume} {83}},\ \bibinfo {pages} {349--404}
  (\bibinfo {year} {2011})}\BibitemShut {NoStop}%
\bibitem [{\citenamefont {S\'emon}\ \emph
  {et~al.}(2014{\natexlab{a}})\citenamefont {S\'emon}, \citenamefont {Yee},
  \citenamefont {Haule},\ and\ \citenamefont {Tremblay}}]{patrickSkipList}%
  \BibitemOpen
  \bibfield  {author} {\bibinfo {author} {\bibfnamefont {P.}~\bibnamefont
  {S\'emon}}, \bibinfo {author} {\bibfnamefont {Chuck-Hou}\ \bibnamefont
  {Yee}}, \bibinfo {author} {\bibfnamefont {Kristjan}\ \bibnamefont {Haule}}, \
  and\ \bibinfo {author} {\bibfnamefont {A.-M.~S.}\ \bibnamefont {Tremblay}},\
  }\bibfield  {title} {\enquote {\bibinfo {title} {{Lazy skip-lists: An
  algorithm for fast hybridization-expansion quantum Monte Carlo}},}\ }\href
  {\doibase 10.1103/PhysRevB.90.075149} {\bibfield  {journal} {\bibinfo
  {journal} {Phys. Rev. B}\ }\textbf {\bibinfo {volume} {90}},\ \bibinfo
  {pages} {075149} (\bibinfo {year} {2014}{\natexlab{a}})}\BibitemShut
  {NoStop}%
\bibitem [{\citenamefont {S\'emon}\ \emph
  {et~al.}(2014{\natexlab{b}})\citenamefont {S\'emon}, \citenamefont {Sordi},\
  and\ \citenamefont {Tremblay}}]{patrickERG}%
  \BibitemOpen
  \bibfield  {author} {\bibinfo {author} {\bibfnamefont {P.}~\bibnamefont
  {S\'emon}}, \bibinfo {author} {\bibfnamefont {G.}~\bibnamefont {Sordi}}, \
  and\ \bibinfo {author} {\bibfnamefont {A.-M.~S.}\ \bibnamefont {Tremblay}},\
  }\bibfield  {title} {\enquote {\bibinfo {title} {Ergodicity of the
  hybridization-expansion monte carlo algorithm for broken-symmetry states},}\
  }\href {\doibase 10.1103/PhysRevB.89.165113} {\bibfield  {journal} {\bibinfo
  {journal} {Phys. Rev. B}\ }\textbf {\bibinfo {volume} {89}},\ \bibinfo
  {pages} {165113} (\bibinfo {year} {2014}{\natexlab{b}})}\BibitemShut
  {NoStop}%
\bibitem [{\citenamefont {Walsh}\ \emph
  {et~al.}(2019{\natexlab{a}})\citenamefont {Walsh}, \citenamefont {S\'emon},
  \citenamefont {Poulin}, \citenamefont {Sordi},\ and\ \citenamefont
  {Tremblay}}]{CaitlinSb}%
  \BibitemOpen
  \bibfield  {author} {\bibinfo {author} {\bibfnamefont {C.}~\bibnamefont
  {Walsh}}, \bibinfo {author} {\bibfnamefont {P.}~\bibnamefont {S\'emon}},
  \bibinfo {author} {\bibfnamefont {D.}~\bibnamefont {Poulin}}, \bibinfo
  {author} {\bibfnamefont {G.}~\bibnamefont {Sordi}}, \ and\ \bibinfo {author}
  {\bibfnamefont {A.-M.~S.}\ \bibnamefont {Tremblay}},\ }\bibfield  {title}
  {\enquote {\bibinfo {title} {{Thermodynamic and information-theoretic
  description of the Mott transition in the two-dimensional Hubbard model}},}\
  }\href {\doibase 10.1103/PhysRevB.99.075122} {\bibfield  {journal} {\bibinfo
  {journal} {Phys. Rev. B}\ }\textbf {\bibinfo {volume} {99}},\ \bibinfo
  {pages} {075122} (\bibinfo {year} {2019}{\natexlab{a}})}\BibitemShut
  {NoStop}%
\bibitem [{\citenamefont {Sordi}\ \emph
  {et~al.}(2012{\natexlab{a}})\citenamefont {Sordi}, \citenamefont {S\'emon},
  \citenamefont {Haule},\ and\ \citenamefont {Tremblay}}]{sshtSC}%
  \BibitemOpen
  \bibfield  {author} {\bibinfo {author} {\bibfnamefont {G.}~\bibnamefont
  {Sordi}}, \bibinfo {author} {\bibfnamefont {P.}~\bibnamefont {S\'emon}},
  \bibinfo {author} {\bibfnamefont {K.}~\bibnamefont {Haule}}, \ and\ \bibinfo
  {author} {\bibfnamefont {A.-M.~S.}\ \bibnamefont {Tremblay}},\ }\bibfield
  {title} {\enquote {\bibinfo {title} {{Strong Coupling Superconductivity,
  Pseudogap, and Mott Transition}},}\ }\href {\doibase
  10.1103/PhysRevLett.108.216401} {\bibfield  {journal} {\bibinfo  {journal}
  {Phys. Rev. Lett.}\ }\textbf {\bibinfo {volume} {108}},\ \bibinfo {pages}
  {216401} (\bibinfo {year} {2012}{\natexlab{a}})}\BibitemShut {NoStop}%
\bibitem [{\citenamefont {Mermin}\ and\ \citenamefont
  {Wagner}(1966)}]{MWtheorem}%
  \BibitemOpen
  \bibfield  {author} {\bibinfo {author} {\bibfnamefont {N.~D.}\ \bibnamefont
  {Mermin}}\ and\ \bibinfo {author} {\bibfnamefont {H.}~\bibnamefont
  {Wagner}},\ }\bibfield  {title} {\enquote {\bibinfo {title} {{Absence of
  Ferromagnetism or Antiferromagnetism in One- or Two-Dimensional Isotropic
  Heisenberg Models}},}\ }\href {\doibase 10.1103/PhysRevLett.17.1133}
  {\bibfield  {journal} {\bibinfo  {journal} {Phys. Rev. Lett.}\ }\textbf
  {\bibinfo {volume} {17}},\ \bibinfo {pages} {1133--1136} (\bibinfo {year}
  {1966})}\BibitemShut {NoStop}%
\bibitem [{\citenamefont {Kosterlitz}\ and\ \citenamefont
  {Thouless}(1973)}]{Kosterlitz:1973}%
  \BibitemOpen
  \bibfield  {author} {\bibinfo {author} {\bibfnamefont {J~M}\ \bibnamefont
  {Kosterlitz}}\ and\ \bibinfo {author} {\bibfnamefont {D~J}\ \bibnamefont
  {Thouless}},\ }\bibfield  {title} {\enquote {\bibinfo {title} {Ordering,
  metastability and phase transitions in two-dimensional systems},}\ }\href
  {\doibase 10.1088/0022-3719/6/7/010} {\bibfield  {journal} {\bibinfo
  {journal} {Journal of Physics C: Solid State Physics}\ }\textbf {\bibinfo
  {volume} {6}},\ \bibinfo {pages} {1181--1203} (\bibinfo {year}
  {1973})}\BibitemShut {NoStop}%
\bibitem [{\citenamefont {Fratino}\ \emph {et~al.}(2016)\citenamefont
  {Fratino}, \citenamefont {S\'emon}, \citenamefont {Sordi},\ and\
  \citenamefont {Tremblay}}]{LorenzoSC}%
  \BibitemOpen
  \bibfield  {author} {\bibinfo {author} {\bibfnamefont {L.}~\bibnamefont
  {Fratino}}, \bibinfo {author} {\bibfnamefont {P.}~\bibnamefont {S\'emon}},
  \bibinfo {author} {\bibfnamefont {G.}~\bibnamefont {Sordi}}, \ and\ \bibinfo
  {author} {\bibfnamefont {A.-M.~S.}\ \bibnamefont {Tremblay}},\ }\bibfield
  {title} {\enquote {\bibinfo {title} {{An organizing principle for
  two-dimensional strongly correlated superconductivity}},}\ }\href {\doibase
  10.1038/srep22715} {\bibfield  {journal} {\bibinfo  {journal} {Sci. Rep.}\
  }\textbf {\bibinfo {volume} {6}},\ \bibinfo {pages} {22715} (\bibinfo {year}
  {2016})}\BibitemShut {NoStop}%
\bibitem [{\citenamefont {Sordi}\ \emph {et~al.}(2013)\citenamefont {Sordi},
  \citenamefont {S\'emon}, \citenamefont {Haule},\ and\ \citenamefont
  {Tremblay}}]{sshtRHO}%
  \BibitemOpen
  \bibfield  {author} {\bibinfo {author} {\bibfnamefont {G.}~\bibnamefont
  {Sordi}}, \bibinfo {author} {\bibfnamefont {P.}~\bibnamefont {S\'emon}},
  \bibinfo {author} {\bibfnamefont {K.}~\bibnamefont {Haule}}, \ and\ \bibinfo
  {author} {\bibfnamefont {A.-M.~S.}\ \bibnamefont {Tremblay}},\ }\bibfield
  {title} {\enquote {\bibinfo {title} {{$c$-axis resistivity, pseudogap,
  superconductivity, and Widom line in doped Mott insulators }},}\ }\href
  {\doibase 10.1103/PhysRevB.87.041101} {\bibfield  {journal} {\bibinfo
  {journal} {Phys. Rev. B}\ }\textbf {\bibinfo {volume} {87}},\ \bibinfo
  {pages} {041101} (\bibinfo {year} {2013})}\BibitemShut {NoStop}%
\bibitem [{\citenamefont {Cyr-Choini\`ere}\ \emph {et~al.}(2018)\citenamefont
  {Cyr-Choini\`ere}, \citenamefont {Daou}, \citenamefont {Lalibert\'e},
  \citenamefont {Collignon}, \citenamefont {Badoux}, \citenamefont {LeBoeuf},
  \citenamefont {Chang}, \citenamefont {Ramshaw}, \citenamefont {Bonn},
  \citenamefont {Hardy}, \citenamefont {Liang}, \citenamefont {Yan},
  \citenamefont {Cheng}, \citenamefont {Zhou}, \citenamefont {Goodenough},
  \citenamefont {Pyon}, \citenamefont {Takayama}, \citenamefont {Takagi},
  \citenamefont {Doiron-Leyraud},\ and\ \citenamefont
  {Taillefer}}]{Olivier:PRB2018}%
  \BibitemOpen
  \bibfield  {author} {\bibinfo {author} {\bibfnamefont {O.}~\bibnamefont
  {Cyr-Choini\`ere}}, \bibinfo {author} {\bibfnamefont {R.}~\bibnamefont
  {Daou}}, \bibinfo {author} {\bibfnamefont {F.}~\bibnamefont {Lalibert\'e}},
  \bibinfo {author} {\bibfnamefont {C.}~\bibnamefont {Collignon}}, \bibinfo
  {author} {\bibfnamefont {S.}~\bibnamefont {Badoux}}, \bibinfo {author}
  {\bibfnamefont {D.}~\bibnamefont {LeBoeuf}}, \bibinfo {author} {\bibfnamefont
  {J.}~\bibnamefont {Chang}}, \bibinfo {author} {\bibfnamefont {B.~J.}\
  \bibnamefont {Ramshaw}}, \bibinfo {author} {\bibfnamefont {D.~A.}\
  \bibnamefont {Bonn}}, \bibinfo {author} {\bibfnamefont {W.~N.}\ \bibnamefont
  {Hardy}}, \bibinfo {author} {\bibfnamefont {R.}~\bibnamefont {Liang}},
  \bibinfo {author} {\bibfnamefont {J.-Q.}\ \bibnamefont {Yan}}, \bibinfo
  {author} {\bibfnamefont {J.-G.}\ \bibnamefont {Cheng}}, \bibinfo {author}
  {\bibfnamefont {J.-S.}\ \bibnamefont {Zhou}}, \bibinfo {author}
  {\bibfnamefont {J.~B.}\ \bibnamefont {Goodenough}}, \bibinfo {author}
  {\bibfnamefont {S.}~\bibnamefont {Pyon}}, \bibinfo {author} {\bibfnamefont
  {T.}~\bibnamefont {Takayama}}, \bibinfo {author} {\bibfnamefont
  {H.}~\bibnamefont {Takagi}}, \bibinfo {author} {\bibfnamefont
  {N.}~\bibnamefont {Doiron-Leyraud}}, \ and\ \bibinfo {author} {\bibfnamefont
  {Louis}\ \bibnamefont {Taillefer}},\ }\bibfield  {title} {\enquote {\bibinfo
  {title} {Pseudogap temperature ${T}^{*}$ of cuprate superconductors from the
  nernst effect},}\ }\href {\doibase 10.1103/PhysRevB.97.064502} {\bibfield
  {journal} {\bibinfo  {journal} {Phys. Rev. B}\ }\textbf {\bibinfo {volume}
  {97}},\ \bibinfo {pages} {064502} (\bibinfo {year} {2018})}\BibitemShut
  {NoStop}%
\bibitem [{\citenamefont {Sordi}\ \emph
  {et~al.}(2012{\natexlab{b}})\citenamefont {Sordi}, \citenamefont {S\'emon},
  \citenamefont {Haule},\ and\ \citenamefont {Tremblay}}]{ssht}%
  \BibitemOpen
  \bibfield  {author} {\bibinfo {author} {\bibfnamefont {G.}~\bibnamefont
  {Sordi}}, \bibinfo {author} {\bibfnamefont {P.}~\bibnamefont {S\'emon}},
  \bibinfo {author} {\bibfnamefont {K.}~\bibnamefont {Haule}}, \ and\ \bibinfo
  {author} {\bibfnamefont {A.-M.~S.}\ \bibnamefont {Tremblay}},\ }\bibfield
  {title} {\enquote {\bibinfo {title} {{Pseudogap temperature as a Widom line
  in doped Mott insulators}},}\ }\href {\doibase doi:10.1038/srep00547}
  {\bibfield  {journal} {\bibinfo  {journal} {Sci. Rep.}\ }\textbf {\bibinfo
  {volume} {2}},\ \bibinfo {pages} {547} (\bibinfo {year}
  {2012}{\natexlab{b}})}\BibitemShut {NoStop}%
\bibitem [{\citenamefont {Sordi}\ \emph {et~al.}(2010)\citenamefont {Sordi},
  \citenamefont {Haule},\ and\ \citenamefont {Tremblay}}]{sht}%
  \BibitemOpen
  \bibfield  {author} {\bibinfo {author} {\bibfnamefont {G.}~\bibnamefont
  {Sordi}}, \bibinfo {author} {\bibfnamefont {K.}~\bibnamefont {Haule}}, \ and\
  \bibinfo {author} {\bibfnamefont {A.-M.~S.}\ \bibnamefont {Tremblay}},\
  }\bibfield  {title} {\enquote {\bibinfo {title} {{Finite Doping Signatures of
  the Mott Transition in the Two-Dimensional Hubbard Model}},}\ }\href
  {\doibase 10.1103/PhysRevLett.104.226402} {\bibfield  {journal} {\bibinfo
  {journal} {Phys. Rev. Lett.}\ }\textbf {\bibinfo {volume} {104}},\ \bibinfo
  {pages} {226402} (\bibinfo {year} {2010})}\BibitemShut {NoStop}%
\bibitem [{\citenamefont {Sordi}\ \emph {et~al.}(2011)\citenamefont {Sordi},
  \citenamefont {Haule},\ and\ \citenamefont {Tremblay}}]{sht2}%
  \BibitemOpen
  \bibfield  {author} {\bibinfo {author} {\bibfnamefont {G.}~\bibnamefont
  {Sordi}}, \bibinfo {author} {\bibfnamefont {K.}~\bibnamefont {Haule}}, \ and\
  \bibinfo {author} {\bibfnamefont {A.-M.~S.}\ \bibnamefont {Tremblay}},\
  }\bibfield  {title} {\enquote {\bibinfo {title} {{Mott physics and
  first-order transition between two metals in the normal-state phase diagram
  of the two-dimensional Hubbard model}},}\ }\href {\doibase
  10.1103/PhysRevB.84.075161} {\bibfield  {journal} {\bibinfo  {journal} {Phys.
  Rev. B}\ }\textbf {\bibinfo {volume} {84}},\ \bibinfo {pages} {075161}
  (\bibinfo {year} {2011})}\BibitemShut {NoStop}%
\bibitem [{\citenamefont {Xu}\ \emph {et~al.}(2005)\citenamefont {Xu},
  \citenamefont {Kumar}, \citenamefont {Buldyrev}, \citenamefont {Chen},
  \citenamefont {Poole}, \citenamefont {Sciortino},\ and\ \citenamefont
  {Stanley}}]{water1}%
  \BibitemOpen
  \bibfield  {author} {\bibinfo {author} {\bibfnamefont {Limei}\ \bibnamefont
  {Xu}}, \bibinfo {author} {\bibfnamefont {Pradeep}\ \bibnamefont {Kumar}},
  \bibinfo {author} {\bibfnamefont {S.~V.}\ \bibnamefont {Buldyrev}}, \bibinfo
  {author} {\bibfnamefont {S.-H.}\ \bibnamefont {Chen}}, \bibinfo {author}
  {\bibfnamefont {P.~H.}\ \bibnamefont {Poole}}, \bibinfo {author}
  {\bibfnamefont {F.}~\bibnamefont {Sciortino}}, \ and\ \bibinfo {author}
  {\bibfnamefont {H.~E.}\ \bibnamefont {Stanley}},\ }\bibfield  {title}
  {\enquote {\bibinfo {title} {{Relation between the Widom line and the dynamic
  crossover in systems with a liquid liquid phase transition}},}\ }\href
  {\doibase 10.1073/pnas.0507870102} {\bibfield  {journal} {\bibinfo  {journal}
  {Proc. Natl. Acad. Sci. USA}\ }\textbf {\bibinfo {volume} {102}},\ \bibinfo
  {pages} {16558--16562} (\bibinfo {year} {2005})}\BibitemShut {NoStop}%
\bibitem [{\citenamefont {Walsh}\ \emph
  {et~al.}(2019{\natexlab{b}})\citenamefont {Walsh}, \citenamefont {S\'emon},
  \citenamefont {Sordi},\ and\ \citenamefont {Tremblay}}]{CaitlinOpalescence}%
  \BibitemOpen
  \bibfield  {author} {\bibinfo {author} {\bibfnamefont {C.}~\bibnamefont
  {Walsh}}, \bibinfo {author} {\bibfnamefont {P.}~\bibnamefont {S\'emon}},
  \bibinfo {author} {\bibfnamefont {G.}~\bibnamefont {Sordi}}, \ and\ \bibinfo
  {author} {\bibfnamefont {A.-M.~S.}\ \bibnamefont {Tremblay}},\ }\bibfield
  {title} {\enquote {\bibinfo {title} {Critical opalescence across the
  doping-driven mott transition in optical lattices of ultracold atoms},}\
  }\href {\doibase 10.1103/PhysRevB.99.165151} {\bibfield  {journal} {\bibinfo
  {journal} {Phys. Rev. B}\ }\textbf {\bibinfo {volume} {99}},\ \bibinfo
  {pages} {165151} (\bibinfo {year} {2019}{\natexlab{b}})}\BibitemShut
  {NoStop}%
\bibitem [{\citenamefont {Sordi}\ \emph {et~al.}(2019)\citenamefont {Sordi},
  \citenamefont {Walsh}, \citenamefont {S\'emon},\ and\ \citenamefont
  {Tremblay}}]{Giovanni:PRBcv}%
  \BibitemOpen
  \bibfield  {author} {\bibinfo {author} {\bibfnamefont {G.}~\bibnamefont
  {Sordi}}, \bibinfo {author} {\bibfnamefont {C.}~\bibnamefont {Walsh}},
  \bibinfo {author} {\bibfnamefont {P.}~\bibnamefont {S\'emon}}, \ and\
  \bibinfo {author} {\bibfnamefont {A.-M.~S.}\ \bibnamefont {Tremblay}},\
  }\bibfield  {title} {\enquote {\bibinfo {title} {Specific heat maximum as a
  signature of mott physics in the two-dimensional hubbard model},}\ }\href
  {\doibase 10.1103/PhysRevB.100.121105} {\bibfield  {journal} {\bibinfo
  {journal} {Phys. Rev. B}\ }\textbf {\bibinfo {volume} {100}},\ \bibinfo
  {pages} {121105} (\bibinfo {year} {2019})}\BibitemShut {NoStop}%
\bibitem [{\citenamefont {{Michon}}\ \emph {et~al.}(2019)\citenamefont
  {{Michon}}, \citenamefont {{Girod}}, \citenamefont {{Badoux}}, \citenamefont
  {{Ka{\v{c}}mar{\v{c}}{\'\i}k}}, \citenamefont {{Ma}}, \citenamefont
  {{Dragomir}}, \citenamefont {{Dabkowska}}, \citenamefont {{Gaulin}},
  \citenamefont {{Zhou}}, \citenamefont {{Pyon}}, \citenamefont {{Takayama}},
  \citenamefont {{Takagi}}, \citenamefont {{Verret}}, \citenamefont
  {{Doiron-Leyraud}}, \citenamefont {{Marcenat}}, \citenamefont {{Taillefer}},\
  and\ \citenamefont {{Klein}}}]{Michon:Cv2018}%
  \BibitemOpen
  \bibfield  {author} {\bibinfo {author} {\bibfnamefont {B.}~\bibnamefont
  {{Michon}}}, \bibinfo {author} {\bibfnamefont {C.}~\bibnamefont {{Girod}}},
  \bibinfo {author} {\bibfnamefont {S.}~\bibnamefont {{Badoux}}}, \bibinfo
  {author} {\bibfnamefont {J.}~\bibnamefont {{Ka{\v{c}}mar{\v{c}}{\'\i}k}}},
  \bibinfo {author} {\bibfnamefont {Q.}~\bibnamefont {{Ma}}}, \bibinfo {author}
  {\bibfnamefont {M.}~\bibnamefont {{Dragomir}}}, \bibinfo {author}
  {\bibfnamefont {H.~A.}\ \bibnamefont {{Dabkowska}}}, \bibinfo {author}
  {\bibfnamefont {B.~D.}\ \bibnamefont {{Gaulin}}}, \bibinfo {author}
  {\bibfnamefont {J.~S.}\ \bibnamefont {{Zhou}}}, \bibinfo {author}
  {\bibfnamefont {S.}~\bibnamefont {{Pyon}}}, \bibinfo {author} {\bibfnamefont
  {T.}~\bibnamefont {{Takayama}}}, \bibinfo {author} {\bibfnamefont
  {H.}~\bibnamefont {{Takagi}}}, \bibinfo {author} {\bibfnamefont
  {S.}~\bibnamefont {{Verret}}}, \bibinfo {author} {\bibfnamefont
  {N.}~\bibnamefont {{Doiron-Leyraud}}}, \bibinfo {author} {\bibfnamefont
  {C.}~\bibnamefont {{Marcenat}}}, \bibinfo {author} {\bibfnamefont
  {L.}~\bibnamefont {{Taillefer}}}, \ and\ \bibinfo {author} {\bibfnamefont
  {T.}~\bibnamefont {{Klein}}},\ }\bibfield  {title} {\enquote {\bibinfo
  {title} {{Thermodynamic signatures of quantum criticality in cuprates}},}\
  }\href {\doibase 10.1038/s41586-019-0932-x} {\bibfield  {journal} {\bibinfo
  {journal} {Nature}\ }\textbf {\bibinfo {volume} {567}},\ \bibinfo {pages}
  {218--222} (\bibinfo {year} {2019})}\BibitemShut {NoStop}%
\bibitem [{\citenamefont {Mikelsons}\ \emph {et~al.}(2009)\citenamefont
  {Mikelsons}, \citenamefont {Khatami}, \citenamefont {Galanakis},
  \citenamefont {Macridin}, \citenamefont {Moreno},\ and\ \citenamefont
  {Jarrell}}]{MikelsonThermodynamics:2009}%
  \BibitemOpen
  \bibfield  {author} {\bibinfo {author} {\bibfnamefont {K.}~\bibnamefont
  {Mikelsons}}, \bibinfo {author} {\bibfnamefont {E.}~\bibnamefont {Khatami}},
  \bibinfo {author} {\bibfnamefont {D.}~\bibnamefont {Galanakis}}, \bibinfo
  {author} {\bibfnamefont {A.}~\bibnamefont {Macridin}}, \bibinfo {author}
  {\bibfnamefont {J.}~\bibnamefont {Moreno}}, \ and\ \bibinfo {author}
  {\bibfnamefont {M.}~\bibnamefont {Jarrell}},\ }\bibfield  {title} {\enquote
  {\bibinfo {title} {Thermodynamics of the quantum critical point at finite
  doping in the two-dimensional hubbard model studied via the dynamical cluster
  approximation},}\ }\href {\doibase 10.1103/PhysRevB.80.140505} {\bibfield
  {journal} {\bibinfo  {journal} {Phys. Rev. B}\ }\textbf {\bibinfo {volume}
  {80}},\ \bibinfo {pages} {140505} (\bibinfo {year} {2009})}\BibitemShut
  {NoStop}%
\bibitem [{\citenamefont {Loram}\ \emph {et~al.}(2001)\citenamefont {Loram},
  \citenamefont {Luo}, \citenamefont {Cooper}, \citenamefont {Liang},\ and\
  \citenamefont {Tallon}}]{LoramJPCS2001}%
  \BibitemOpen
  \bibfield  {author} {\bibinfo {author} {\bibfnamefont {J.W.}\ \bibnamefont
  {Loram}}, \bibinfo {author} {\bibfnamefont {J.}~\bibnamefont {Luo}}, \bibinfo
  {author} {\bibfnamefont {J.R.}\ \bibnamefont {Cooper}}, \bibinfo {author}
  {\bibfnamefont {W.Y.}\ \bibnamefont {Liang}}, \ and\ \bibinfo {author}
  {\bibfnamefont {J.L.}\ \bibnamefont {Tallon}},\ }\bibfield  {title} {\enquote
  {\bibinfo {title} {Evidence on the pseudogap and condensate from the
  electronic specific heat},}\ }\href {\doibase
  https://doi.org/10.1016/S0022-3697(00)00101-3} {\bibfield  {journal}
  {\bibinfo  {journal} {Journal of Physics and Chemistry of Solids}\ }\textbf
  {\bibinfo {volume} {62}},\ \bibinfo {pages} {59--64} (\bibinfo {year}
  {2001})}\BibitemShut {NoStop}%
\bibitem [{\citenamefont {Tallon}\ \emph {et~al.}(2004)\citenamefont {Tallon},
  \citenamefont {Benseman}, \citenamefont {Williams},\ and\ \citenamefont
  {Loram}}]{Tallon:2004}%
  \BibitemOpen
  \bibfield  {author} {\bibinfo {author} {\bibfnamefont {J.L.}\ \bibnamefont
  {Tallon}}, \bibinfo {author} {\bibfnamefont {T.}~\bibnamefont {Benseman}},
  \bibinfo {author} {\bibfnamefont {G.V.M.}\ \bibnamefont {Williams}}, \ and\
  \bibinfo {author} {\bibfnamefont {J.W.}\ \bibnamefont {Loram}},\ }\bibfield
  {title} {\enquote {\bibinfo {title} {The phase diagram of high-tc
  superconductors},}\ }\href {\doibase
  https://doi.org/10.1016/j.physc.2004.07.014} {\bibfield  {journal} {\bibinfo
  {journal} {Physica C: Superconductivity}\ }\textbf {\bibinfo {volume}
  {415}},\ \bibinfo {pages} {9--14} (\bibinfo {year} {2004})}\BibitemShut
  {NoStop}%
\bibitem [{\citenamefont {Zanardi}(2002)}]{zanardi2002}%
  \BibitemOpen
  \bibfield  {author} {\bibinfo {author} {\bibfnamefont {Paolo}\ \bibnamefont
  {Zanardi}},\ }\bibfield  {title} {\enquote {\bibinfo {title} {Quantum
  entanglement in fermionic lattices},}\ }\href {\doibase
  10.1103/PhysRevA.65.042101} {\bibfield  {journal} {\bibinfo  {journal} {Phys.
  Rev. A}\ }\textbf {\bibinfo {volume} {65}},\ \bibinfo {pages} {042101}
  (\bibinfo {year} {2002})}\BibitemShut {NoStop}%
\bibitem [{\citenamefont {Cardy}\ and\ \citenamefont
  {Herzog}(2014)}]{Cardy:2017}%
  \BibitemOpen
  \bibfield  {author} {\bibinfo {author} {\bibfnamefont {John}\ \bibnamefont
  {Cardy}}\ and\ \bibinfo {author} {\bibfnamefont {Christopher~P.}\
  \bibnamefont {Herzog}},\ }\bibfield  {title} {\enquote {\bibinfo {title}
  {Universal thermal corrections to single interval entanglement entropy for
  two dimensional conformal field theories},}\ }\href {\doibase
  10.1103/PhysRevLett.112.171603} {\bibfield  {journal} {\bibinfo  {journal}
  {Phys. Rev. Lett.}\ }\textbf {\bibinfo {volume} {112}},\ \bibinfo {pages}
  {171603} (\bibinfo {year} {2014})}\BibitemShut {NoStop}%
\bibitem [{\citenamefont {Vedral}(2004)}]{Vedral:T2004}%
  \BibitemOpen
  \bibfield  {author} {\bibinfo {author} {\bibfnamefont {Vlatko}\ \bibnamefont
  {Vedral}},\ }\bibfield  {title} {\enquote {\bibinfo {title} {High-temperature
  macroscopic entanglement},}\ }\href
  {http://stacks.iop.org/1367-2630/6/i=1/a=102} {\bibfield  {journal} {\bibinfo
   {journal} {New Journal of Physics}\ }\textbf {\bibinfo {volume} {6}},\
  \bibinfo {pages} {102} (\bibinfo {year} {2004})}\BibitemShut {NoStop}%
\bibitem [{\citenamefont {Anders}\ and\ \citenamefont
  {Vedral}(2007)}]{Anders:2007}%
  \BibitemOpen
  \bibfield  {author} {\bibinfo {author} {\bibfnamefont {Janet}\ \bibnamefont
  {Anders}}\ and\ \bibinfo {author} {\bibfnamefont {Vlatko}\ \bibnamefont
  {Vedral}},\ }\bibfield  {title} {\enquote {\bibinfo {title} {Macroscopic
  entanglement and phase transitions},}\ }\href {\doibase
  10.1007/s11080-007-9034-6} {\bibfield  {journal} {\bibinfo  {journal} {Open
  Systems \& Information Dynamics}\ }\textbf {\bibinfo {volume} {14}},\
  \bibinfo {pages} {1--16} (\bibinfo {year} {2007})}\BibitemShut {NoStop}%
\bibitem [{\citenamefont {Walsh}\ \emph
  {et~al.}(2019{\natexlab{c}})\citenamefont {Walsh}, \citenamefont {S\'emon},
  \citenamefont {Poulin}, \citenamefont {Sordi},\ and\ \citenamefont
  {Tremblay}}]{Caitlin:PRL2019}%
  \BibitemOpen
  \bibfield  {author} {\bibinfo {author} {\bibfnamefont {C.}~\bibnamefont
  {Walsh}}, \bibinfo {author} {\bibfnamefont {P.}~\bibnamefont {S\'emon}},
  \bibinfo {author} {\bibfnamefont {D.}~\bibnamefont {Poulin}}, \bibinfo
  {author} {\bibfnamefont {G.}~\bibnamefont {Sordi}}, \ and\ \bibinfo {author}
  {\bibfnamefont {A.-M.~S.}\ \bibnamefont {Tremblay}},\ }\bibfield  {title}
  {\enquote {\bibinfo {title} {Local entanglement entropy and mutual
  information across the mott transition in the two-dimensional hubbard
  model},}\ }\href {\doibase 10.1103/PhysRevLett.122.067203} {\bibfield
  {journal} {\bibinfo  {journal} {Phys. Rev. Lett.}\ }\textbf {\bibinfo
  {volume} {122}},\ \bibinfo {pages} {067203} (\bibinfo {year}
  {2019}{\natexlab{c}})}\BibitemShut {NoStop}%
\bibitem [{\citenamefont {Gu}\ \emph {et~al.}(2004)\citenamefont {Gu},
  \citenamefont {Deng}, \citenamefont {Li},\ and\ \citenamefont
  {Lin}}]{Gu:2004}%
  \BibitemOpen
  \bibfield  {author} {\bibinfo {author} {\bibfnamefont {Shi-Jian}\
  \bibnamefont {Gu}}, \bibinfo {author} {\bibfnamefont {Shu-Sa}\ \bibnamefont
  {Deng}}, \bibinfo {author} {\bibfnamefont {You-Quan}\ \bibnamefont {Li}}, \
  and\ \bibinfo {author} {\bibfnamefont {Hai-Qing}\ \bibnamefont {Lin}},\
  }\bibfield  {title} {\enquote {\bibinfo {title} {{Entanglement and Quantum
  Phase Transition in the Extended Hubbard Model}},}\ }\href {\doibase
  10.1103/PhysRevLett.93.086402} {\bibfield  {journal} {\bibinfo  {journal}
  {Phys. Rev. Lett.}\ }\textbf {\bibinfo {volume} {93}},\ \bibinfo {pages}
  {086402} (\bibinfo {year} {2004})}\BibitemShut {NoStop}%
\bibitem [{\citenamefont {Deng}\ \emph {et~al.}(2006)\citenamefont {Deng},
  \citenamefont {Gu},\ and\ \citenamefont {Lin}}]{dengPRB2006}%
  \BibitemOpen
  \bibfield  {author} {\bibinfo {author} {\bibfnamefont {Shu-Sa}\ \bibnamefont
  {Deng}}, \bibinfo {author} {\bibfnamefont {Shi-Jian}\ \bibnamefont {Gu}}, \
  and\ \bibinfo {author} {\bibfnamefont {Hai-Qing}\ \bibnamefont {Lin}},\
  }\bibfield  {title} {\enquote {\bibinfo {title} {Block-block entanglement and
  quantum phase transitions in the one-dimensional extended hubbard model},}\
  }\href {\doibase 10.1103/PhysRevB.74.045103} {\bibfield  {journal} {\bibinfo
  {journal} {Phys. Rev. B}\ }\textbf {\bibinfo {volume} {74}},\ \bibinfo
  {pages} {045103} (\bibinfo {year} {2006})}\BibitemShut {NoStop}%
\bibitem [{\citenamefont {Maier}\ \emph {et~al.}(2004)\citenamefont {Maier},
  \citenamefont {Jarrell}, \citenamefont {Macridin},\ and\ \citenamefont
  {Slezak}}]{maierENERGY}%
  \BibitemOpen
  \bibfield  {author} {\bibinfo {author} {\bibfnamefont {Th.~A.}\ \bibnamefont
  {Maier}}, \bibinfo {author} {\bibfnamefont {M.}~\bibnamefont {Jarrell}},
  \bibinfo {author} {\bibfnamefont {A.}~\bibnamefont {Macridin}}, \ and\
  \bibinfo {author} {\bibfnamefont {C.}~\bibnamefont {Slezak}},\ }\bibfield
  {title} {\enquote {\bibinfo {title} {Kinetic energy driven pairing in cuprate
  superconductors},}\ }\href {\doibase 10.1103/PhysRevLett.92.027005}
  {\bibfield  {journal} {\bibinfo  {journal} {Phys. Rev. Lett.}\ }\textbf
  {\bibinfo {volume} {92}},\ \bibinfo {pages} {027005} (\bibinfo {year}
  {2004})}\BibitemShut {NoStop}%
\bibitem [{\citenamefont {Carbone}\ \emph {et~al.}(2006)\citenamefont
  {Carbone}, \citenamefont {Kuzmenko}, \citenamefont {Molegraaf}, \citenamefont
  {van Heumen}, \citenamefont {Lukovac}, \citenamefont {Marsiglio},
  \citenamefont {van~der Marel}, \citenamefont {Haule}, \citenamefont
  {Kotliar}, \citenamefont {Berger}, \citenamefont {Courjault}, \citenamefont
  {Kes},\ and\ \citenamefont {Li}}]{carbone2006}%
  \BibitemOpen
  \bibfield  {author} {\bibinfo {author} {\bibfnamefont {F.}~\bibnamefont
  {Carbone}}, \bibinfo {author} {\bibfnamefont {A.~B.}\ \bibnamefont
  {Kuzmenko}}, \bibinfo {author} {\bibfnamefont {H.~J.~A.}\ \bibnamefont
  {Molegraaf}}, \bibinfo {author} {\bibfnamefont {E.}~\bibnamefont {van
  Heumen}}, \bibinfo {author} {\bibfnamefont {V.}~\bibnamefont {Lukovac}},
  \bibinfo {author} {\bibfnamefont {F.}~\bibnamefont {Marsiglio}}, \bibinfo
  {author} {\bibfnamefont {D.}~\bibnamefont {van~der Marel}}, \bibinfo {author}
  {\bibfnamefont {K.}~\bibnamefont {Haule}}, \bibinfo {author} {\bibfnamefont
  {G.}~\bibnamefont {Kotliar}}, \bibinfo {author} {\bibfnamefont
  {H.}~\bibnamefont {Berger}}, \bibinfo {author} {\bibfnamefont
  {S.}~\bibnamefont {Courjault}}, \bibinfo {author} {\bibfnamefont {P.~H.}\
  \bibnamefont {Kes}}, \ and\ \bibinfo {author} {\bibfnamefont
  {M.}~\bibnamefont {Li}},\ }\bibfield  {title} {\enquote {\bibinfo {title}
  {{Doping dependence of the redistribution of optical spectral weight in
  ${\mathrm{Bi}}_{2}{\mathrm{Sr}}_{2}{\mathrm{CaCu}}_{2}{\mathrm{O}}_{8+\ensuremath{\delta}}$}},}\
  }\href {\doibase 10.1103/PhysRevB.74.064510} {\bibfield  {journal} {\bibinfo
  {journal} {Phys. Rev. B}\ }\textbf {\bibinfo {volume} {74}},\ \bibinfo
  {pages} {064510} (\bibinfo {year} {2006})}\BibitemShut {NoStop}%
\bibitem [{\citenamefont {Gull}\ and\ \citenamefont
  {Millis}(2012)}]{millisENERGY}%
  \BibitemOpen
  \bibfield  {author} {\bibinfo {author} {\bibfnamefont {E.}~\bibnamefont
  {Gull}}\ and\ \bibinfo {author} {\bibfnamefont {A.~J.}\ \bibnamefont
  {Millis}},\ }\bibfield  {title} {\enquote {\bibinfo {title} {Energetics of
  superconductivity in the two-dimensional hubbard model},}\ }\href {\doibase
  10.1103/PhysRevB.86.241106} {\bibfield  {journal} {\bibinfo  {journal} {Phys.
  Rev. B}\ }\textbf {\bibinfo {volume} {86}},\ \bibinfo {pages} {241106}
  (\bibinfo {year} {2012})}\BibitemShut {NoStop}%
\bibitem [{\citenamefont {Groisman}\ \emph {et~al.}(2005)\citenamefont
  {Groisman}, \citenamefont {Popescu},\ and\ \citenamefont
  {Winter}}]{groisman2005}%
  \BibitemOpen
  \bibfield  {author} {\bibinfo {author} {\bibfnamefont {Berry}\ \bibnamefont
  {Groisman}}, \bibinfo {author} {\bibfnamefont {Sandu}\ \bibnamefont
  {Popescu}}, \ and\ \bibinfo {author} {\bibfnamefont {Andreas}\ \bibnamefont
  {Winter}},\ }\bibfield  {title} {\enquote {\bibinfo {title} {Quantum,
  classical, and total amount of correlations in a quantum state},}\ }\href
  {\doibase 10.1103/PhysRevA.72.032317} {\bibfield  {journal} {\bibinfo
  {journal} {Phys. Rev. A}\ }\textbf {\bibinfo {volume} {72}},\ \bibinfo
  {pages} {032317} (\bibinfo {year} {2005})}\BibitemShut {NoStop}%
\bibitem [{\citenamefont {Wolf}\ \emph {et~al.}(2008)\citenamefont {Wolf},
  \citenamefont {Verstraete}, \citenamefont {Hastings},\ and\ \citenamefont
  {Cirac}}]{wolfPRL2008}%
  \BibitemOpen
  \bibfield  {author} {\bibinfo {author} {\bibfnamefont {Michael~M.}\
  \bibnamefont {Wolf}}, \bibinfo {author} {\bibfnamefont {Frank}\ \bibnamefont
  {Verstraete}}, \bibinfo {author} {\bibfnamefont {Matthew~B.}\ \bibnamefont
  {Hastings}}, \ and\ \bibinfo {author} {\bibfnamefont {J.~Ignacio}\
  \bibnamefont {Cirac}},\ }\bibfield  {title} {\enquote {\bibinfo {title} {Area
  laws in quantum systems: Mutual information and correlations},}\ }\href
  {\doibase 10.1103/PhysRevLett.100.070502} {\bibfield  {journal} {\bibinfo
  {journal} {Phys. Rev. Lett.}\ }\textbf {\bibinfo {volume} {100}},\ \bibinfo
  {pages} {070502} (\bibinfo {year} {2008})}\BibitemShut {NoStop}%
\bibitem [{\citenamefont {Vidal}\ and\ \citenamefont
  {Werner}(2002)}]{vidal:PRA2002}%
  \BibitemOpen
  \bibfield  {author} {\bibinfo {author} {\bibfnamefont {G.}~\bibnamefont
  {Vidal}}\ and\ \bibinfo {author} {\bibfnamefont {R.~F.}\ \bibnamefont
  {Werner}},\ }\bibfield  {title} {\enquote {\bibinfo {title} {Computable
  measure of entanglement},}\ }\href {\doibase 10.1103/PhysRevA.65.032314}
  {\bibfield  {journal} {\bibinfo  {journal} {Phys. Rev. A}\ }\textbf {\bibinfo
  {volume} {65}},\ \bibinfo {pages} {032314} (\bibinfo {year}
  {2002})}\BibitemShut {NoStop}%
\bibitem [{\citenamefont {Shastry}(2009)}]{Shastry:2009}%
  \BibitemOpen
  \bibfield  {author} {\bibinfo {author} {\bibfnamefont {B~Sriram}\
  \bibnamefont {Shastry}},\ }\bibfield  {title} {\enquote {\bibinfo {title}
  {Electrothermal transport coefficients at finite frequencies},}\ }\href@noop
  {} {\bibfield  {journal} {\bibinfo  {journal} {Reports on Progress in
  Physics}\ }\textbf {\bibinfo {volume} {72}},\ \bibinfo {pages} {016501}
  (\bibinfo {year} {2009})}\BibitemShut {NoStop}%
\bibitem [{\citenamefont {Honma}\ and\ \citenamefont {Hor}(2008)}]{Honma:2008}%
  \BibitemOpen
  \bibfield  {author} {\bibinfo {author} {\bibfnamefont {T.}~\bibnamefont
  {Honma}}\ and\ \bibinfo {author} {\bibfnamefont {P.~H.}\ \bibnamefont
  {Hor}},\ }\bibfield  {title} {\enquote {\bibinfo {title} {Unified electronic
  phase diagram for hole-doped high- $t_{c}$ cuprates},}\ }\href {\doibase
  10.1103/PhysRevB.77.184520} {\bibfield  {journal} {\bibinfo  {journal} {Phys.
  Rev. B}\ }\textbf {\bibinfo {volume} {77}},\ \bibinfo {pages} {184520}
  (\bibinfo {year} {2008})}\BibitemShut {NoStop}%
\bibitem [{\citenamefont {Chakraborty}\ \emph {et~al.}(2010)\citenamefont
  {Chakraborty}, \citenamefont {Galanakis},\ and\ \citenamefont
  {Phillips}}]{ShilaPRB2010}%
  \BibitemOpen
  \bibfield  {author} {\bibinfo {author} {\bibfnamefont {Shiladitya}\
  \bibnamefont {Chakraborty}}, \bibinfo {author} {\bibfnamefont {Dimitrios}\
  \bibnamefont {Galanakis}}, \ and\ \bibinfo {author} {\bibfnamefont {Philip}\
  \bibnamefont {Phillips}},\ }\bibfield  {title} {\enquote {\bibinfo {title}
  {Emergence of particle-hole symmetry near optimal doping in high-temperature
  copper oxide superconductors},}\ }\href {\doibase 10.1103/PhysRevB.82.214503}
  {\bibfield  {journal} {\bibinfo  {journal} {Phys. Rev. B}\ }\textbf {\bibinfo
  {volume} {82}},\ \bibinfo {pages} {214503} (\bibinfo {year}
  {2010})}\BibitemShut {NoStop}%
\bibitem [{\citenamefont {Collignon}\ \emph {et~al.}(2021)\citenamefont
  {Collignon}, \citenamefont {Ataei}, \citenamefont {Gourgout}, \citenamefont
  {Badoux}, \citenamefont {Lizaire}, \citenamefont {Legros}, \citenamefont
  {Licciardello}, \citenamefont {Wiedmann}, \citenamefont {Yan}, \citenamefont
  {Zhou}, \citenamefont {Ma}, \citenamefont {Gaulin}, \citenamefont
  {Doiron-Leyraud},\ and\ \citenamefont {Taillefer}}]{CollignonPRB2021}%
  \BibitemOpen
  \bibfield  {author} {\bibinfo {author} {\bibfnamefont {C.}~\bibnamefont
  {Collignon}}, \bibinfo {author} {\bibfnamefont {A.}~\bibnamefont {Ataei}},
  \bibinfo {author} {\bibfnamefont {A.}~\bibnamefont {Gourgout}}, \bibinfo
  {author} {\bibfnamefont {S.}~\bibnamefont {Badoux}}, \bibinfo {author}
  {\bibfnamefont {M.}~\bibnamefont {Lizaire}}, \bibinfo {author} {\bibfnamefont
  {A.}~\bibnamefont {Legros}}, \bibinfo {author} {\bibfnamefont
  {S.}~\bibnamefont {Licciardello}}, \bibinfo {author} {\bibfnamefont
  {S.}~\bibnamefont {Wiedmann}}, \bibinfo {author} {\bibfnamefont {J.-Q.}\
  \bibnamefont {Yan}}, \bibinfo {author} {\bibfnamefont {J.-S.}\ \bibnamefont
  {Zhou}}, \bibinfo {author} {\bibfnamefont {Q.}~\bibnamefont {Ma}}, \bibinfo
  {author} {\bibfnamefont {B.~D.}\ \bibnamefont {Gaulin}}, \bibinfo {author}
  {\bibfnamefont {Nicolas}\ \bibnamefont {Doiron-Leyraud}}, \ and\ \bibinfo
  {author} {\bibfnamefont {Louis}\ \bibnamefont {Taillefer}},\ }\bibfield
  {title} {\enquote {\bibinfo {title} {{Thermopower across the phase diagram of
  the cuprate
  ${\mathrm{La}}_{1.6\ensuremath{-}x}{\mathrm{Nd}}_{0.4}{\mathrm{Sr}}_{x}{\mathrm{CuO}}_{4}$:
  Signatures of the pseudogap and charge density wave phases}},}\ }\href
  {\doibase 10.1103/PhysRevB.103.155102} {\bibfield  {journal} {\bibinfo
  {journal} {Phys. Rev. B}\ }\textbf {\bibinfo {volume} {103}},\ \bibinfo
  {pages} {155102} (\bibinfo {year} {2021})}\BibitemShut {NoStop}%
\bibitem [{\citenamefont {S\'en\'echal}\ and\ \citenamefont
  {Tremblay}(2004)}]{senechalPRL2004}%
  \BibitemOpen
  \bibfield  {author} {\bibinfo {author} {\bibfnamefont {David}\ \bibnamefont
  {S\'en\'echal}}\ and\ \bibinfo {author} {\bibfnamefont {A.-M.~S.}\
  \bibnamefont {Tremblay}},\ }\bibfield  {title} {\enquote {\bibinfo {title}
  {Hot spots and pseudogaps for hole- and electron-doped high-temperature
  superconductors},}\ }\href {\doibase 10.1103/PhysRevLett.92.126401}
  {\bibfield  {journal} {\bibinfo  {journal} {Phys. Rev. Lett.}\ }\textbf
  {\bibinfo {volume} {92}},\ \bibinfo {pages} {126401} (\bibinfo {year}
  {2004})}\BibitemShut {NoStop}%
\bibitem [{\citenamefont {Wald}\ \emph {et~al.}(2020)\citenamefont {Wald},
  \citenamefont {Arias},\ and\ \citenamefont {Alba}}]{Wald:JSM2020}%
  \BibitemOpen
  \bibfield  {author} {\bibinfo {author} {\bibfnamefont {Sascha}\ \bibnamefont
  {Wald}}, \bibinfo {author} {\bibfnamefont {Ra{\'{u}}l}\ \bibnamefont
  {Arias}}, \ and\ \bibinfo {author} {\bibfnamefont {Vincenzo}\ \bibnamefont
  {Alba}},\ }\bibfield  {title} {\enquote {\bibinfo {title} {Entanglement and
  classical fluctuations at finite-temperature critical points},}\ }\href
  {\doibase 10.1088/1742-5468/ab6b19} {\bibfield  {journal} {\bibinfo
  {journal} {Journal of Statistical Mechanics: Theory and Experiment}\ }\textbf
  {\bibinfo {volume} {2020}},\ \bibinfo {pages} {033105} (\bibinfo {year}
  {2020})}\BibitemShut {NoStop}%
\bibitem [{\citenamefont {Walsh}\ \emph {et~al.}(Deposited 27 April
  2021)\citenamefont {Walsh}, \citenamefont {Charlebois}, \citenamefont
  {Sémon}, \citenamefont {Sordi},\ and\ \citenamefont {Tremblay}}]{dataSCENT}%
  \BibitemOpen
  \bibfield  {author} {\bibinfo {author} {\bibfnamefont {C.}~\bibnamefont
  {Walsh}}, \bibinfo {author} {\bibfnamefont {M.}~\bibnamefont {Charlebois}},
  \bibinfo {author} {\bibfnamefont {P.}~\bibnamefont {Sémon}}, \bibinfo
  {author} {\bibfnamefont {G.}~\bibnamefont {Sordi}}, \ and\ \bibinfo {author}
  {\bibfnamefont {A.-M.S.}\ \bibnamefont {Tremblay}},\ }\href {\doibase
  10.17605/OSF.IO/A8B4Y} {\enquote {\bibinfo {title} {{Data associated with
  "Information-theoretic measures of superconductivity in a two-dimensional
  doped Mott insulator", Open Science Framework (OSF)}},}\ } (\bibinfo {year}
  {Deposited 27 April 2021})\BibitemShut {NoStop}%
\end{thebibliography}

%

\onecolumngrid
\clearpage
\setcounter{figure}{0}
\setcounter{section}{0}
\makeatletter 
\renewcommand{\thefigure}{S\@arabic\c@figure}

\begin{center}

{\bf Supplementary Information for} \\
\vspace{0.05cm} 

{\bf Information-theoretic measures of superconductivity in a two-dimensional doped Mott insulator} \\
\vspace{0.05cm}

{C. Walsh, M. Charlebois, P. S\'emon, G. Sordi, and A.-M. S. Tremblay}

\end{center}

\begin{figure}[b!]
\centering
\includegraphics[width=0.8\textwidth]{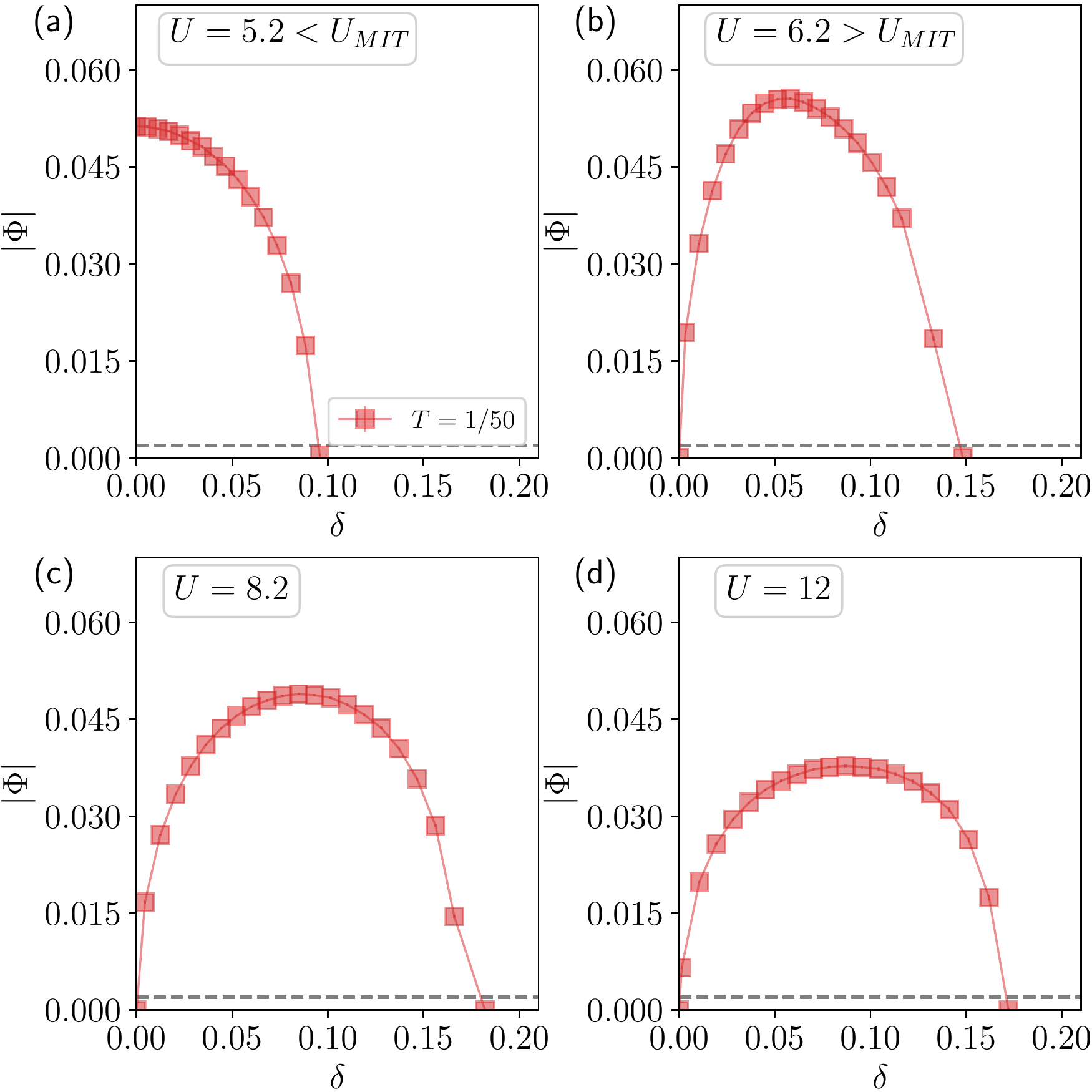}
\caption{Superconducting order parameter $|\Phi|$ as a function of doping $\delta$ for different values of the interaction strength $U$ at $T=1/50$. Superconductivity is indicated by a nonzero value of $\Phi$. Operationally, here we consider the system to be superconducting when $|\Phi|>0.002$ (see dashed horizontal line). (a) $U=5.2$, (b) $U=6.2$, (c) $U=8.2$, (d) $U=12$. }
\end{figure}

\begin{figure}
\centering
\includegraphics[width=0.8\textwidth]{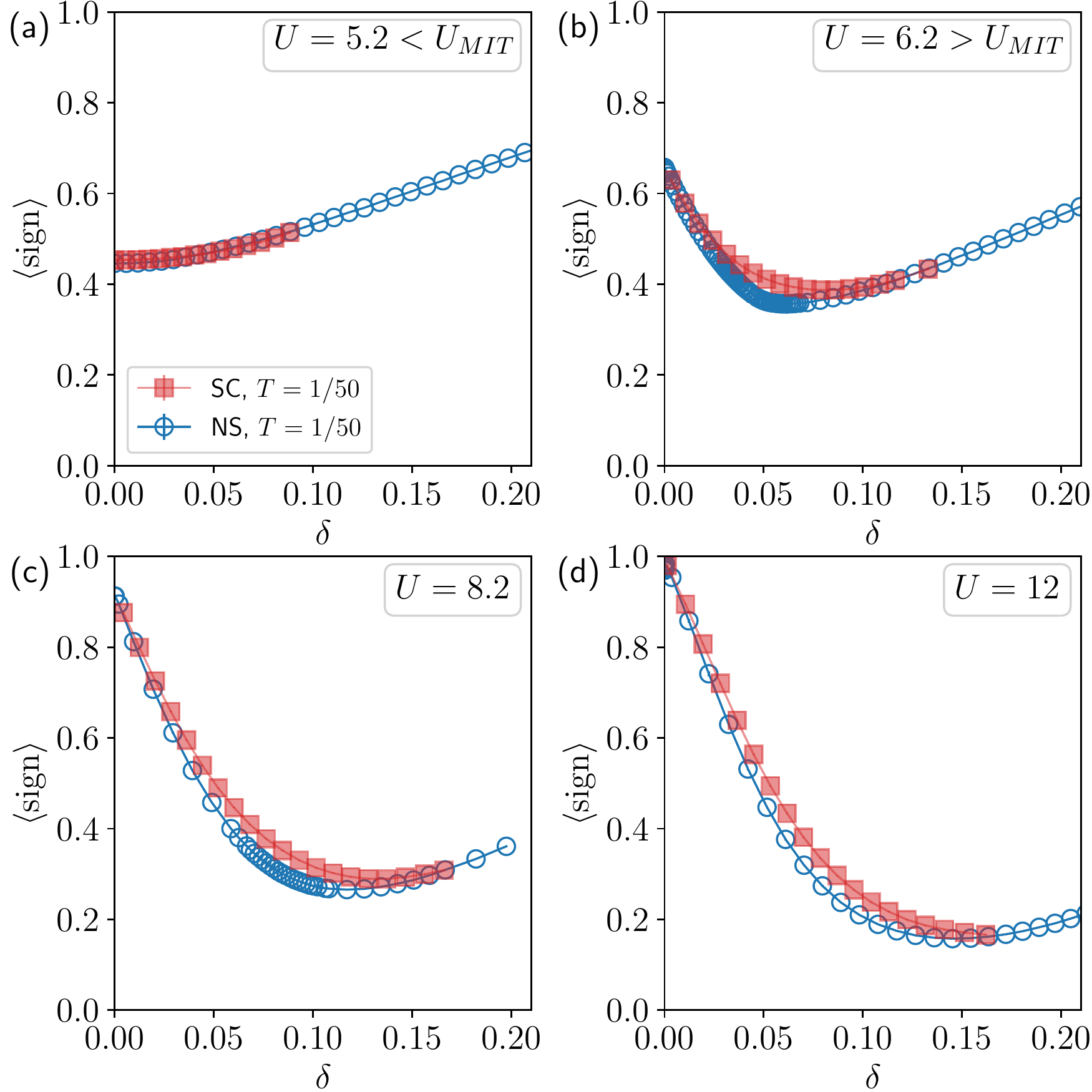}
\caption{Average Monte Carlo sign as a function of doping $\delta$ for different values of the interaction strength $U$ at $T=1/50$, for both the normal state (open blue circles) and superconducting state (filled red squares). (a) $U=5.2$, (b) $U=6.2$, (c) $U=8.2$, (d) $U=12$. }
\end{figure}

\begin{figure}
\centering
\includegraphics[width=\textwidth]{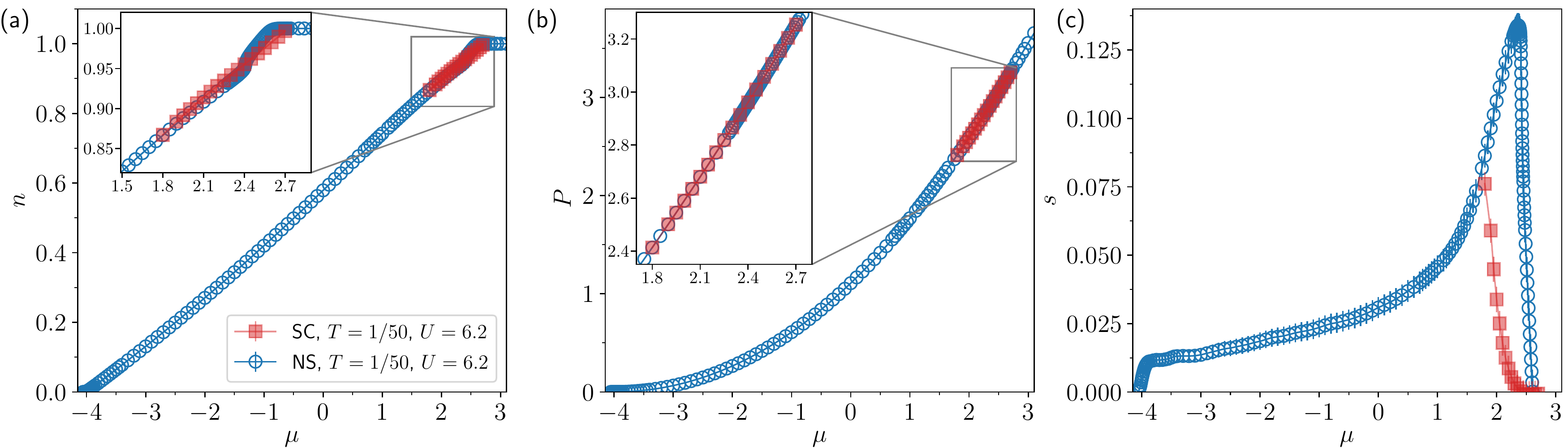}
\caption{Occupation $n$, pressure $P$, and thermodynamic entropy $s$ versus $\mu$ at $T=1/50$ and $U=6.2$ for both the normal state (open blue circles) and superconducting state (filled red squares). These panels demonstrate the steps in the procedure for calculating $s$ as described in the ``Materials and Methods'' section of the main text.}
\end{figure}

\end{document}